\documentclass[aps,prb,twocolumn,nofootinbib,superscriptaddress]{revtex4-2}
\usepackage[dvipdfmx]{graphicx}
\usepackage{dcolumn}
\usepackage{times}
\usepackage{color}
\usepackage{bm}
\usepackage{braket}
\usepackage{physics}
\usepackage[version=4]{mhchem}
\usepackage{amsmath, amssymb}
\usepackage[colorlinks=true, citecolor=blue]{hyperref}
\usepackage{multirow}
\newcommand{\nn}{\nonumber}

\hypersetup{
setpagesize=false,
 bookmarksnumbered=true,%
 bookmarksopen=true,%
 colorlinks=true,%
 linkcolor=black,
 citecolor=blue,
}

\begin{document}
\preprint{APS/123-QED}
\title{
Magnetic penetration depth in topological superconductors: Effect of Majorana surface states and application for UTe\texorpdfstring{$_2$}{2}
}
\author{Kazuki Akuzawa}
\affiliation{Department of Materials Engineering Science, The University of Osaka, Toyonaka,  560-8531, Japan}
\author{Jushin Tei}
\email{tei@blade.mp.es.osaka-u.ac.jp}
\affiliation{Department of Materials Engineering Science, The University of Osaka, Toyonaka,  560-8531, Japan}
\author{Ryoi Ohashi}
\affiliation{Department of Physics, Kyushu University, Fukuoka, 819-0395, Japan}
\author{Satoshi Fujimoto}
\email{fuji@mp.es.osaka-u.ac.jp}
\affiliation{Department of Materials Engineering Science, The University of Osaka, Toyonaka, 560-8531, Japan}
\affiliation{Center for Quantum Information and Quantum Biology, The University of Osaka, Toyonaka 560-8531, Japan}
\affiliation{Center for Spintronics Research Network, Graduate School of Engineering Science, The University of Osaka,  Toyonaka, 560-8531, Japan}
\author{Takeshi Mizushima}
\email{mizushima@mp.es.osaka-u.ac.jp}
\affiliation{Department of Materials Engineering Science, The University of Osaka, Toyonaka, 560-8531, Japan}
\date{\today}

\begin{abstract}
In this study, we examine how orbital degrees of freedom and Majorana surface states influence the magnetic penetration depth in the superconductor UTe$_2$. Using a two-orbital model, we analyze pairing states belonging to the irreducible representations of the $D_{2h}$ crystal symmetry: $A_u$, $B_{1u}$, $B_{2u}$, and $B_{3u}$. For bulk nodal states such as $B_{2u}$, we find that the penetration depth for screening currents along the antinodal direction and the cylindrical axis scales as $T^2$, in strong contrast to the conventional $T^4$ law. This behavior originates from quasiparticles near the point nodes contributing to the interorbital paramagnetic current. We further show that Majorana surface states can dominate the low-temperature response. The fully gapped $A_u$ state hosts Majorana cones, which produce a $T^3$ dependence of the penetration depth when the ratio of penetration depth to coherence length ($\kappa$) is small. {In contrast, the other pairing states exhibit Majorana Fermi arcs: the exponent is $n=2$ along the dispersive direction, while along the dispersionless direction it depends on whether the arcs terminate at endpoints. The exponent $n=2$ in the dispersive direction is robust, while it in the dispersionless direction relies on the presence or absence of the endpoints of the arcs and deviates from $n=2$ when endpoints are absent. Our results demonstrate that penetration-depth measurements provide a direct probe of Majorana surface states in low-$\kappa$ superconductors. For larger $\kappa$, the surface contribution becomes negligible and the temperature dependence is governed by bulk quasiparticles.}
\end{abstract}

\maketitle

\section{Introduction}
\begin{figure}[t]
    \centering
    \includegraphics[width=\linewidth]{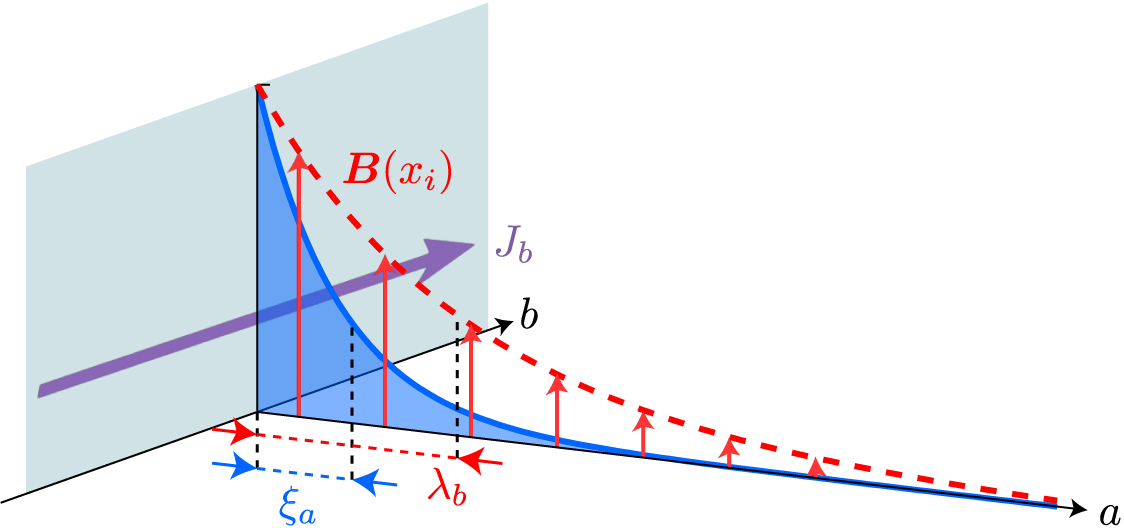} 
    \caption{An external magnetic field is screened by the superconducting current, $J_b$, resulting in an exponential decay of the magnetic field over the characteristic length scale of the magnetic penetration depth, $\lambda_b$.
    The blue curve illustrates the probability density of the Majorana surface states, which are localized within the coherence length $\xi_a$ from the boundary.}
    \label{Fig:setup}
\end{figure}

Since superconductivity was first observed in \ce{UTe2}, this heavy-fermion compound has attracted considerable interest~\cite{ran19,aoki19,aok22,lew23}. It is a leading candidate for materials that host spin-triplet unconventional superconductivity. Several experimental observations support spin-triplet superconductivity, including extremely high upper critical fields in all directions~\cite{ran_2019a,aoki_2020,knebel_2020} and reentrant superconductivity at high magnetic fields~\cite{ran_2019a,miyake19,knafo19}. Furthermore, the reduction in NMR Knight shift for all directions is compatible with a fully gapped spin-triplet pairing state~\cite{matsumura_2023}.
However, identifying the superconducting gap symmetry of UTe$_2$ still remains a crucial unresolved challenge, even under low magnetic fields and ambient pressure. 

\begin{figure}[t]
    \centering
    \includegraphics[width=\linewidth]{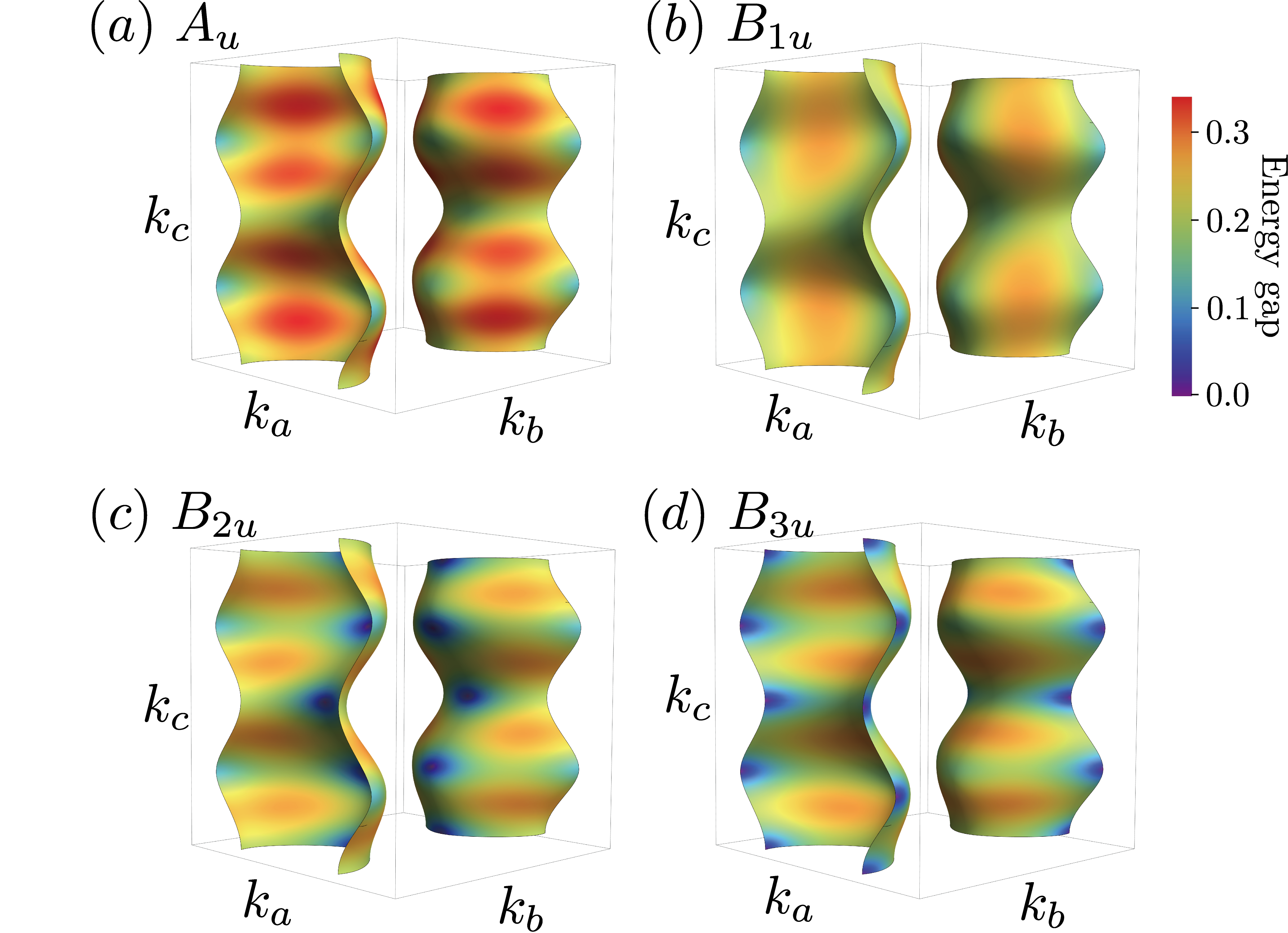} 
    \caption{Cylindrical Fermi surfaces and gap structures for (a)~$A_u$, (b)~$B_{1u}$, (c)~$B_{2u}$, and (d)~$B_{3u}$ states, respectively, where $k_a{a},~k_b{b}\in[-\pi,\pi]$ and $k_c{c}\in [-2\pi,2\pi]$.
    The color represents the magnitude of the superconducting gap.
    The $A_u$ and $B_{1u}$ states are fully gapped, whereas the $B_{2u}$ and $B_{3u}$ states have point nodes.}
    \label{Fig:Fermi}
\end{figure}

Possible spin-triplet pairing states in \ce{UTe2} are classified in terms of four irreducible representations of the $D_{2h}$ point group symmetry, the $A_{u}$, $B_{1u}$, $B_{2u}$, and $B_{3u}$ states. Besides the gap symmetry, information about the electronic band structure is also essential for understanding the unconventional superconductivity of \ce{UTe2}. The recent observation of the de Haas–van Alphen oscillations supports the existence of cylindrical Fermi surfaces along the $k_c$ direction~\cite{aoki2022First}. For such Fermi surfaces, the $A_{u}$ and $B_{1u}$ states are fully gapped, while the other two states have nodal points along the $k_a$ or $k_b$ direction~[see Fig.~\ref{Fig:Fermi}]. However, the pairing symmetry suggested by recent experiments remains controversial and has yet to be established. For instance, the spontaneous breaking of time-reversal symmetry has been reported in samples with $T_{\rm c}=1.6~{\rm K}$~\cite{jiao_2020_Chiral,hayes_2021_Multicomponent}, while the recent remarkable improvement in sample quality increases the superconducting transition temperature at zero fields from $T_{\rm c}=1.6~{\rm K}$ to $2.1~{\rm K}$~\cite{sakai_2022_Single,rosa_2022_Single}, and Kerr effect and zero-field muon spin relaxation measurements on high-quality samples have shown no evidence of broken time-reversal symmetry~\cite{ajeesh_2023,aza23}. The nodal structures of the superconducting gap may be identified experimentally through the power-law behaviors in the temperature dependence of thermal and magnetic measurements. In another instance, thermal conductivity measurements yield conflicting results for the gap symmetries: One group indicates a fully gapped $A_u$ state~\cite{suetsugu_2024_Fully}, while another presents data indicating the existence of point nodes~\cite{hayes_2024}. Furthermore, measurements of specific heat and magnetic penetration depth indicate the presence of point nodes, suggesting a nonunitary $B_{3u}+iA_{u}$ state with spontaneously broken time-reversal symmetry~\cite{ish23}. As a result, the irreducible representation and gap symmetry have yet to be determined.

Among several experiments, we here focus on magnetic penetration depth measurements~\cite{met19,bae21,ish23}. According to the theory based on bulk gap structures~\cite{einzel86,gross86}, in superconductors with nodal points, the magnetic field penetration depth of the Meissner current along the nodal (antinodal) direction behaves as $T^2$ ($T^4$). Recent experiment~\cite{ish23} reports $T^2$ dependence of superfluid density in all directions, suggesting a nonunitary pairing state with point nodes. However, in the case with time-reversal symmetry, all possible pairing states in \ce{UTe2} belong to topological phases protected by crystalline symmetries, and Majorana quasiparticles residing on the surfaces are responsible for anomalous magnetic response and tunneling phenomena~\cite{tei2023possible,ohashi2024anisotropic,hyeok24,li25,gu25,chris25,tei25}. Therefore, it is natural to consider that these surface states could influence the penetration depth. Indeed, {it has been predicted that Majorana surface states lead to a $T^3$ power-law suppression of the superfluid mass current flowing in a narrow channel of superfluid $^3$He-B~\cite{hao13,sauls22arxiv}.} In a fully gapped superconductor, which belongs to topological class DIII, the Majorana cone has been shown to be responsible for the paramagnetic effect, resulting in the $T^3$-dependence of magnetic penetration depth~\cite{wu2020power,sauls22}. In UTe$_2$ with cylindrical Fermi surfaces, the $A_{u}$ state hosts Majorana cones, while Majorana Fermi arcs protected by magnetic mirror symmetries appear on the particular surface orientations in the other pairing states~\cite{tei2023possible,ohashi2024anisotropic,tei25}. It is essential to clarify how differences in the dispersion of topological surface states affect the screening current and the temperature dependence of magnetic penetration depth.

In this paper, we investigate the magnetic penetration depth in the superconductor \ce{UTe2} with a special focus on orbital degrees of freedom and Majorana surface states. Here, we use an effective model involving two orbital degrees of freedom of $f$-electrons, which arise from two uranium atoms in a unit cell of the crystal structure. We first consider bulk nodal superconductors without surface states and find that the magnetic penetration depth of the screening current along the antinodal direction exhibits the temperature dependence of $T^2$, which significantly deviates from $T^4$ expected from conventional theory~\cite{einzel86,gross86}. This anomalous exponent is attributed to the quasiparticle contributions around the point nodes to the inter-orbital paramagnetic current. 

{We demonstrate that Majorana surface states can leave clear and measurable signatures in the temperature dependence of the magnetic penetration depth when the zero-temperature penetration depth becomes comparable to the coherence length. Based on a power-counting analysis of a single-band model, we show that the exponent of the surface paramagnetic current is controlled by the dimensionality of the Majorana spectrum: a Majorana cone produces a $T^3$ dependence, whereas Majorana arcs generically lead to $T^2$. Importantly, the analysis reveals that whether the arcs terminate in the surface Brillouin zone qualitatively affects the electromagnetic response—the presence or absence of arc endpoints strongly modifies the exponent of the screening current along the dispersionless direction at intermediate temperatures. Using an effective tight-binding model for UTe$_2$, we show that these general principles account for the penetration-depth exponents of candidate odd-parity pairing states. Our results suggest that penetration-depth measurements in low-$\kappa$ superconductors provide a direct route to detecting Majorana surface states, while in the large-$\kappa$ limit the response is dominated by bulk quasiparticles.}

The organization of this paper is as follows. In Sec.~\ref{sec:topology}, we review the topological superconductivity of UTe$_2$ and the existence of Majorana surface states. We also introduce the model Hamiltonian involving two orbital $f$-electrons. In Sec.~\ref{sec:formulation}, the basic formulation for magnetic penetration depth in superconductors with open boundary conditions. We present the numerical results of the magnetic penetration depth in Secs.~\ref{sec:bulk} and \ref{sec:majorana}. In Sec.~\ref{sec:bulk}, we clarify the effects of inter-orbital transitions on the penetration depth in bulk superconductors without surface states, while the effect of Majorana surface states is discussed in Sec.~\ref{sec:majorana}. Section~\ref{sec:conclusion} is devoted to the conclusion and discussion.

\section{topological superconducting states in UTe\texorpdfstring{$_2$}{2}}
\label{sec:topology}

In this section, we introduce the basic properties of superconductivity in UTe$_2$ and present a model Hamiltonian that captures its essential features.
We also briefly summarize the Majorana surface states. In this paper, we set $\hbar = k_{\rm B}=1$. 

\subsection{BdG Hamiltonian of UTe\texorpdfstring{$_2$}{2}}
\label{sec:bdg}
We start with the Bogoliubov-de Gennes (BdG) Hamiltonian,
\begin{eqnarray}
    \mathcal{H} = \frac{1}{2}\sum_{\bm{k},\alpha,\alpha^{\prime}}\bm{c}_{\bm{k}\alpha}^{\dagger}
    H_{\rm BdG}(\bm{k})\bm{c}_{\bm{k},\alpha^{\prime}},
\end{eqnarray}
where 
\begin{eqnarray}
    H_{\rm BdG}(\bm{k}) = 
    \begin{pmatrix}
        H_{\rm N}(\bm{k}) & \Delta(\bm{k}) \\
        \Delta^{\dagger}(\bm{k}) & -H^{\rm tr}_{\rm N}(-\bm{k})
    \end{pmatrix}.
    \label{eq:Hbdg}
\end{eqnarray}
The BdG Hamiltonian consists of a normal-state part $H_{\rm N}$ and the pair potential $\Delta$, where $A^{\rm tr}$ is the transpose of a matrix $A$. We have also defined $\bm{c}_{\bm{k}\alpha} =  ( c_{\bm{k}\alpha}, c^{\dagger}_{-\bm{k}\alpha})^{\rm tr}$, where $c_{\bm{k}\alpha}$ and $c^{\dagger}_{\bm{k}\alpha}$ are the annihilation and creation operators of electrons with crystal momentum $\bm{k}$ and orbital index $\alpha$, respectively.

UTe$_2$ belongs to the space group $Immm$ (No.71, $D^{25}_{2h}$) and has a body-centered orthorhombic crystal structure.
Within each unit cell, two uranium atoms form dimers aligned along the $c$ axis. Regarding the normal states,
de Haas-van Alphen (dHvA) experiments have revealed two kinds of cylindrical Fermi surface---electron surfaces and hole surfaces~\cite{aoki2022First,eaton2024quasi}.
Despite their cylindrical shape, the observed weak anisotropy in the transport coefficients along the $c$-axis indicates that these Fermi surfaces are quasi-2D but slightly warped along the $c$-axis~\cite{eo2022caxis}.
The electron sheets primarily comprise $5f$ electrons from U sites, whereas the hole sheets include $6d$ electrons at U sites and $5p$ electrons at Te sites~\cite{xu19,ishizuka_2021_Periodic}. Here, we assume that strong Coulomb repulsion among $f$-electrons induces magnetic fluctuations that mediate spin-triplet odd-parity pairing. To capture the Fermi surface geometry of the electron band, therefore, we adopt a simple two-orbital model that focuses only on the $f$-electrons of the uranium dimers~\cite{shishidou2021topological,tei2023possible},
\begin{eqnarray}
    \label{eq:Normal}
    H_{\rm N}(\bm{k}) = (\epsilon_0(\bm{k}) - \mu)\tau_0 + f_x(\bm{k})\tau_x + f_y(\bm{k})\tau_y ,
\end{eqnarray}
where
\begin{gather}
\label{eq:epsilon}
    \epsilon_0(\bm{k}) = 2t_1\cos k_aa + 2t_2 \cos k_bb + 2t_3\cos k_cc,\\
    f_x(\bm{k}) = t_4 + t_5\cos(k_aa/2)\cos(k_bb/2)\cos(k_cc/2), \\
    f_y(\bm{k}) = t_6\cos(k_aa/2)\cos(k_bb/2)\sin(k_cc/2). 
\end{gather}
Here, $a$, $b$, and $c$ are the lattice constants for each crystallographic axis, and $\boldsymbol{\tau}$ is the Pauli matrix acting on the uranium-orbital space.
The terms proportional to $\tau_0$ represent intra-orbital kinetic processes, 
while those proportional to $\tau_x$ and $\tau_y$ describe inter-orbital hopping.
To reproduce the quasi-2D Fermi surface shown in Fig.~\ref{Fig:Fermi}, we set parameters as follows:
$\mu=2.8W$,~$t_1=0.75W$,~$t_2=-1.0W$,~$t_3=-0.3W$,~$t_4=-1.4W$,~$t_5=1.3W$,~$t_6=1.3W$, where we introduce the energy scale $W$, which approximately corresponds to the bandwidth.
Note that this model only reproduces the electron-like Fermi surface observed in UTe$_2$.
The hole-like Fermi surfaces are obtained by rotating the electron-like Fermi surfaces by 90 degrees around the $c$-axis, and the effects of the hole-like Fermi surfaces are considered in analogy with the results for the electronic surfaces.

Next, we consider the gap function, assuming inter-orbital spin-triplet Cooper pairing,
which is given by
\begin{eqnarray}
\label{eq:delta}
    \Delta({\bm k}) = \bm{d}({\bm k})\cdot\boldsymbol{\sigma}i\sigma_y\tau_x,
\end{eqnarray}
where the vectorial order parameter, ${\bm d}(\bm{k})$, characterizes the spin structure of the spin-triplet pairing, and $\bm{\sigma}$ are the Pauli matrices in spin space. {Here we consider only interorbital pairing states to simplify the analysis of the surface response. However, including intraorbital pairing does not alter the  qualitative features of the results, such as the topological protection of the surface ABS~\cite{tei2023possible}.}
In the point group $D_{2h}$, which corresponds to the space group symmetry of UTe$_2$,
there are four possible one-dimensional irreducible representations (IRs) for odd-parity superconductivity: $A_u$, $B_{1u}$, $B_{2u}$, and $B_{3u}$.
The corresponding $d$-vector for each IR is given by
\begin{gather}
    \bm{d}_{A_u} = \Delta_0(T)(\sin k_aa,~\sin k_bb,~\sin k_cc), \\
    \bm{d}_{B_{1u}} = \Delta_0(T)(\sin k_bb,~\sin k_aa,~0), \\
    \bm{d}_{B_{2u}} = \Delta_0(T)(\sin k_cc,~0,~\sin k_aa), \\
    \bm{d}_{B_{3u}} = \Delta_0(T)(0,~\sin k_cc,~\sin k_bb).
\end{gather}
The temperature dependence of the gap function is assumed to follow that in the BCS theory as  
$\Delta_0(T) = \Delta_0\tanh(1.74\sqrt{T_{\rm c}/T -1})$ with $\Delta_0=0.2W$. {The choice $\Delta_0=0.2W$ simply sets the energy scale, as the topologically protected Majorana arc dispersion is insensitive to its magnitude. We also set $\Delta_0$ to be uniform in space. Although self-consistency would alter the gap profile near the surface over the scale of the coherence length, the low-energy structure of the Majorana surface states and the resulting power-law behavior remain unchanged.}

Figure~\ref{Fig:Fermi} illustrates the nodal structures for all IRs.
Both $A_u$ and $B_{1u}$ are fully gapped. This is because $B_{1u}$ has nodal points along the $c$ axis, which opens up the Fermi surface.
$B_{2u}$ and $B_{3u}$ are point-nodal superconductors. For cylindrical Fermi surfaces, several nodal points appear: $k_a=0$, $k_c =  0, \pm \pi /{c}, \pm 2\pi/{c}$ in $B_{2u}$ and $k_b=\pi{/b}$, $k_c = 0, \pm \pi{/c}, \pm 2\pi{/c}$ in $B_{3u}$. Not all nodes are necessarily protected by symmetry. 
{In the present work, we focus on superconductors in symmetry class DIII. Owing to the coexistence of time-reversal and particle-hole symmetries, the system possesses chiral symmetry. Each nodal point is therefore protected by chiral symmetry and characterized by a topological winding number. Indeed, the nodes and the associated Majorana arcs remain stable even when additional symmetry-allowed terms are introduced, such as an orbital-singlet pairing component or higher-harmonic terms like $\sin(k_a) \sin(k_b) \sin(k_c)$~\cite{tei2023possible,tei25}.}

\subsection{Majorana surface states}
\begin{figure}[t]
    \centering
    \includegraphics[width=\linewidth]{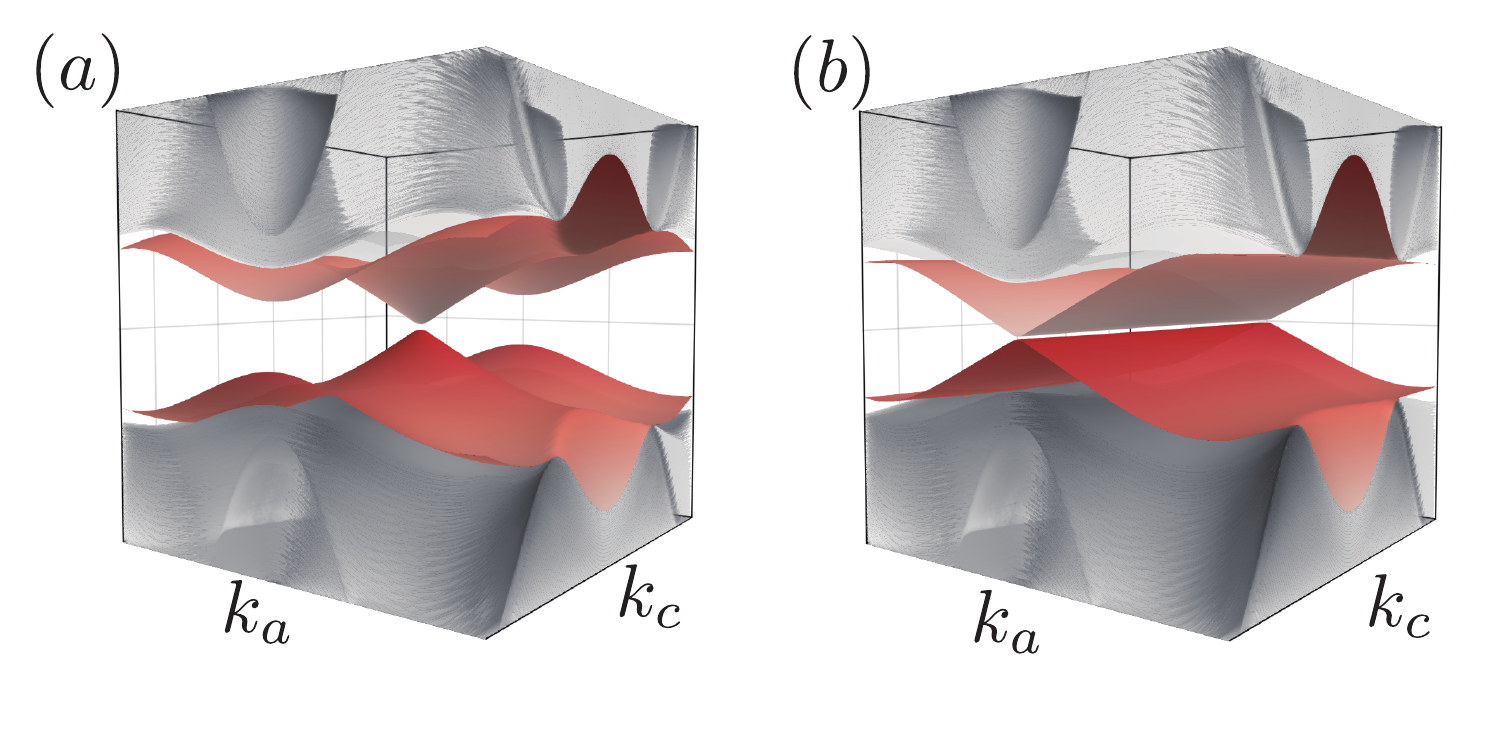} 
    \caption{(a)~Typical structure of energy spectra for Majorana cone systems.
    The result is a slab geometry of the $A_u$ state with open $(010)$ surfaces.
    (a)~Typical structure of energy spectra for Majorana Fermi arc systems.
    The result is a slab geometry of the $B_{1u}$ state with open $(010)$ surfaces.}  % 全体のキャプション
    \label{Fig:Majo}
\end{figure}
\begin{figure}[t]
    \centering
    \includegraphics[width=\linewidth]{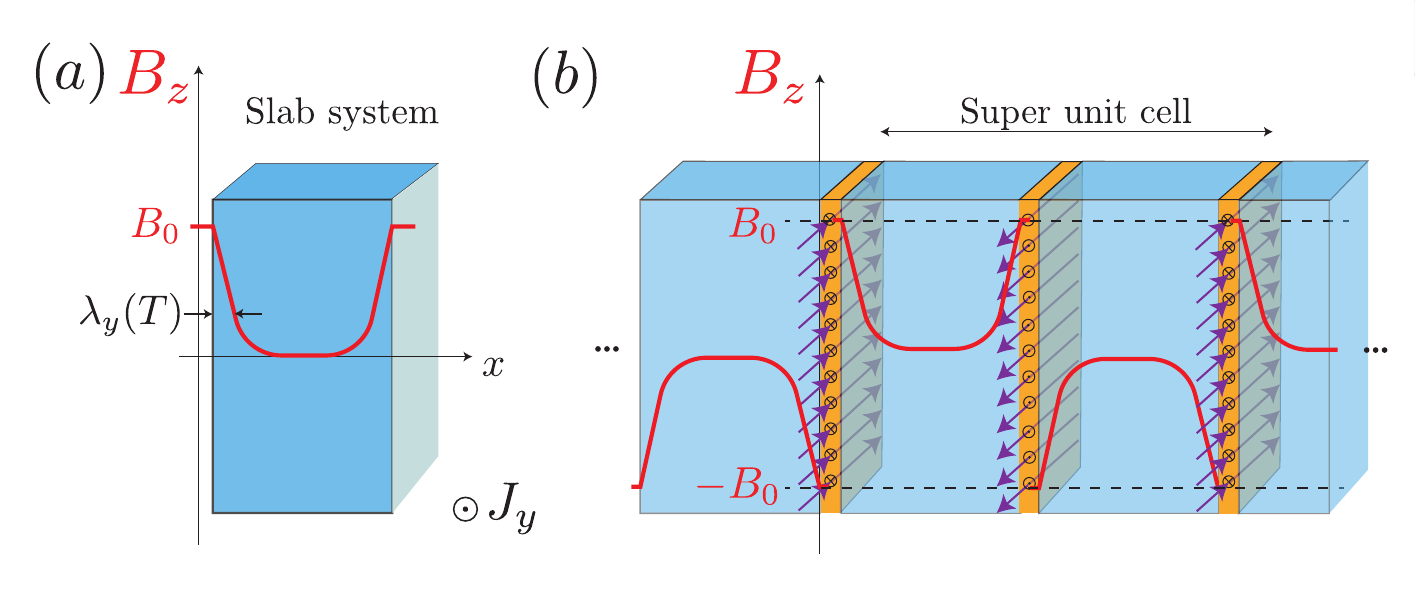} 
    \caption{(a)~Slab geometry with open surfaces perpendicular to the $x$-axis.
    An external magnetic field applied in the $z$-direction is screened by a Meissner current flowing along the $y$-axis.
    (b)~Schematic illustration of the method of electrical mirror imaging.
    A planar current flowing in a thin slab reproduces the external magnetic field.}  
    \label{Fig:mirror}
\end{figure}

Let us consider a slab geometry with open surfaces perpendicular to the $x$-axis, as shown in Fig.~\ref{Fig:mirror}(a).
In this system, the BdG Hamiltonian in Eq.~\eqref{eq:Hbdg} can be recast into
\small
\begin{align}
    H_{\rm BdG}^{\rm lattice}(\bm{k}_{\parallel}) = 
    \begin{pmatrix}
        {H}_0(\bm{k}_{\parallel}) & {T}(\bm{k}_{\parallel}) & 0 & \cdots & 0 \\
        {T}^{\dagger}(\bm{k}_{\parallel}) & {H}_0(\bm{k}_{\parallel}) & {T}(\bm{k}_{\parallel}) &  \cdots & 0\\
        0 & {T}^{\dagger}(\bm{k}_{\parallel}) & {H}_0(\bm{k}_{\parallel})  & \ddots & \vdots \\
        \vdots & \vdots & \ddots &\ddots & {T}(\bm{k}_{\parallel}) \\
        0 & 0 & \cdots&{T}^{\dagger}(\bm{k}_{\parallel}) & {H}_0(\bm{k}_{\parallel})
    \end{pmatrix},
    \label{eq:Hbdg_lattice}
\end{align}
\normalsize
where ${\bm k}_{\parallel}$ represents the momenta parallel to the surface. The matrix is expressed in the basis of lattice sites along the $x$-direction.
The submatrices, ${H}_0$ and ${T}$, are the $8\times 8$ matrices of the on-site Hamiltonian and the nearest-neighbor hopping terms in the spin, orbital, and Nambu bases, respectively. For simplicity, we neglect hopping processes beyond nearest neighbors.

Previous studies~\cite{tei2023possible,ohashi2024anisotropic} examined  Majorana surface states by numerically diagonalizing the BdG Hamiltonian and analyzing the associated topological invariants.
When Fermi surfaces are quasi-two-dimensional and cylindrical in shape, 
the crystalline symmetries play crucial roles in determining the topological nature of the superconducting state.
For example, using the crystalline symmetry operation $U$, one can define a crystalline chiral operator as
\begin{eqnarray}
    \Gamma_{U} = e^{i\phi}U\Theta C,
\end{eqnarray} 
which enables the definition of a 1D winding number as a topological invariant.
Here, $e^{i\phi}$ is the phase factor chosen to satisfy the condition $\Gamma_U^2=1$, $\Theta$ is the time-reversal operator,
and $C$ is the particle-hole exchange operator.
Table~\ref{table:topo} summarizes the surface Majorana states and corresponding crystalline chiral operators for each surface and each IR. The surface states in the $A_{u}$ state are protected by the twofold rotation symmetry about the surface normal axis $C_{\mu}$, while flat Majorana arcs appear in the other IRs as a consequence of the mirror reflection symmetry $M_{\mu\nu}$ with respect to the $\mu$-$\nu$ plane ($\mu,\nu=a,b,c$).

\begin{table}[t!]
    \caption{Summary of Majorana surface states and associated crystalline chiral operators for all IRs, where $C_{\mu}$ and $M_{\mu\nu}$ are the twofold rotation about the $\mu$-axis and the mirror reflection with respect to the $\mu$-$\nu$ plane, respectively ($\mu,\nu=a,b,c$).}
    \centering
    \begin{ruledtabular}
    \begin{tabular}{cccc}
      IR  & (100)            & (010)            & (001)            \\ \hline
    $A_u$  & Majorana cone  & Majorana cone  & ---              \\
    & $\Gamma_{C_a}$ & $\Gamma_{C_b}$ \\
    $B_{1u}$ & Flat Majorana arc & Flat Majorana arc & ---              \\
    & $\Gamma_{M_{ca}}$ & $\Gamma_{M_{bc}}$ \\
    $B_{2u}$ & Flat Majorana arc & ---              & Flat Majorana arc \\
    & $\Gamma_{M_{ab}}$ & & $\Gamma_{M_{bc}}$ \\
    $B_{3u}$ & ---              & Flat Majorana arc & ---              \\
    & & $\Gamma_{M_{ab}}$ &  \\ 
    \end{tabular}
    \end{ruledtabular}
    \label{table:topo}
\end{table}

Figures~\ref{Fig:Majo}(a) and \ref{Fig:Majo}(b) show the low-lying quasiparticle excitation spectra for the $(010)$ surface of the $A_u$ and $B_{1u}$ states, respectively. A Majorana cone appears on the surface of the $A_{u}$ state, while a Majorana arc is formed on the surface of the $B_{1u}$ state. In the latter case, the surface state is dispersionless along the nodal direction and dispersive along the other direction in the surface Brillouin zone.
The $B_{2u}$ and $B_{3u}$ states also exhibit Majorana arcs, similar to those in the $B_{1u}$ state~\cite{tei2023possible,ohashi2024anisotropic}.
We note that the bulk excitations in the $B_{2u}$ and $B_{3u}$ states are gapless due to the existence of nodal points on the cylindrical Fermi surfaces.

Majorana surface states are localized within the coherence length, defined in BCS theory as $\xi_i = \hbar (v_F)_i/\Delta$, $i = a, b, c$, where $v_F$ is the Fermi velocity.
In this work, we approximate the coherence length as  $\xi_a/a = t_1/\Delta_0 \approx 5$, $ \xi_b/b = t_2/\Delta_0 \approx 4$,
and $\xi_c/c = t_3/\Delta_0 \approx 2$.

\section{Formulation for magnetic penetration depth}
\label{sec:formulation}
In this section, we describe the method for calculating the magnetic penetration depth in a tight-binding model for slab geometry with open boundary conditions.
First, we calculate the current response to a static electromagnetic field using linear response theory. Then, by plugging the current response into Maxwell's equation and solving it under appropriate boundary conditions, we derive the spatial profile of the magnetic field penetrating into the superconductor.

\subsection{Electric current response}
We consider a slab geometry in Cartesian coordinates $(x,y,z)$ with open surfaces along the $x$-direction as shown in Fig.~\ref{Fig:mirror}(a).
The lattice constants along the $x$, $y$, and $z$ directions are denoted by $a$, $b$, and $c$, respectively.
The momenta parallel to the surface are denoted by $\bm{k}_{\parallel} = (k_y,k_z)$. 
The number of lattice sites along the $x$-axis is denoted by $L$, with positions labeled as $x_i = 1,\dots,L$.
In the presence of an external magnetic field $B_z$ along the $z$-axis, a Meissner current $J_y$ flows along the $y$-direction to screen the magnetic field inside the superconductor.
The external magnetic field is described by the vector potential $A_y(x_i)$ as
\begin{eqnarray}
    \label{eq:B}
    B_z(x_i) = (A_y(x_i+1) - A_y(x_i) )/a.
\end{eqnarray}
The vector potential couples to the Hamiltonian via minimal coupling, implemented as
\begin{eqnarray}
    k_y \rightarrow k_y - eA_y(x_{i})/\hbar c_0,
    \label{eq:minimal}
\end{eqnarray}
where $c_0$ is the speed of light.

The current operator is defined as the functional derivative of the Hamiltonian with respect to the vector potential:
\begin{eqnarray}
    \hat{J}_{y}(x_i) = - c_0\frac{\delta \mathcal{H}}{\delta A_{y}(x_i)}.
\end{eqnarray}
Note that the off-diagonal components in Nambu space break gauge invariance and therefore do not contribute to the current operator.
To this end, we introduce the Hamiltonian that retains only the diagonal components in Nambu space.
\begin{align}
    \tilde{\mathcal{H}}_{\rm lattice} = & \sum_{\bm{k}_{\parallel},x_i,\alpha,\alpha'}
    \bm{c}^{\dagger}_{\bm{k}_{\parallel},x_i,\alpha}[{\tilde{H}}_0(\bm{k}_{\parallel})]_{\alpha\alpha'}\bm{c}_{\bm{k}_{\parallel},x_i,\alpha'} \nn \\
    & + \bm{c}^{\dagger}_{\bm{k}_{\parallel},x_i,\alpha}[{\tilde{T}}(\bm{k}_{\parallel})]_{\alpha\alpha'}\bm{c}_{\bm{k}_{\parallel},x_i+1,\alpha'} + \rm{h.c},
\end{align}
where $\tilde{H}_0({\bm k}_{\parallel})$ and $\tilde{T}({\bm k}_{\parallel})$ are the diagonal parts of onsite and nearest-neighbor BdG hopping terms in the matrix of Eq.~\eqref{eq:Hbdg_lattice}, respectively.
To simplify the description, we omit the orbital index from now on. We introduce the vector potential into the Hamiltonian through minimal coupling in Eq.~\eqref{eq:minimal} and expand the Hamiltonian to the linear order in $A_y(x_i)$. By taking the functional derivative of the resulting Hamiltonian, we express the current operator as
\begin{eqnarray}
    \hat{J}_y(x_i) = \hat{J}^{\rm para}(x_i) - \frac{1}{c_0}\hat{K}^{\rm dia}(x_i)A_y(x_i),
\end{eqnarray}
where $\hat{J}^{\rm para}(x_i)$ is the paramagnetic current
\begin{align}
    &\hat{J}^{\rm para}(x_i) = \frac{e}{\hbar}\sum_{\bm{k}_{\parallel}}
    \bigg[\bm{c}^{\dagger}_{\bm{k}_{\parallel},x_i}[\partial_{k_y}\tilde{H}_0(\bm{k}_{\parallel})]\bm{c}_{\bm{k}_{\parallel},x_{i}} \nn \\
    &+ \frac{1}{2} [\partial_{k_y}\tilde{T}(\bm{k}_{\parallel})]\Big\{ \bm{c}^{\dagger}_{\bm{k}_{\parallel},x_i}\bm{c}_{\bm{k}_{\parallel},x_{i}+1} + \bm{c}^{\dagger}_{\bm{k}_{\parallel},x_i-1}\bm{c}_{\bm{k}_{\parallel},x_{i}} \Big\} \nn \\
    &+ \frac{1}{2} [\partial_{k_y}\tilde{T}^{\dagger}(\bm{k}_{\parallel})]\Big\{ \bm{c}^{\dagger}_{\bm{k}_{\parallel},x_i+1}\bm{c}_{\bm{k}_{\parallel},x_{i}} + \bm{c}^{\dagger}_{\bm{k}_{\parallel},x_i}\bm{c}_{\bm{k}_{\parallel},x_{i}-1} \Big\} \bigg],
\end{align}
and $\tilde{K}^{\rm dia}(x_i)$ is the diamagnetic kernel
\begin{align}
    &\hat{K}^{\rm dia}(x_i) = -\frac{e^2}{\hbar^2}\sum_{\bm{k}_{\parallel}}
    \bigg[
    \bm{c}^{\dagger}_{\bm{k}_{\parallel},x_i}[\partial^2_{k_y}\tilde{H}_0(\bm{k}_{\parallel})]\bm{c}_{\bm{k}_{\parallel},x_{i}} \nn \\
    &+ \frac{1}{2}\sum_{\bm{k}_{\parallel}} [\partial^2_{k_y}\tilde{T}(\bm{k}_{\parallel})]\Big\{ \bm{c}^{\dagger}_{\bm{k}_{\parallel},x_i}\bm{c}_{\bm{k}_{\parallel},x_{i}+1} + \bm{c}^{\dagger}_{\bm{k}_{\parallel},x_i-1}\bm{c}_{\bm{k}_{\parallel},x_{i}} \Big\} \nn \\
    &+ \frac{1}{2}\sum_{\bm{k}_{\parallel}} [\partial^2_{k_y}\tilde{T}^{\dagger}(\bm{k}_{\parallel})]\Big\{ \bm{c}^{\dagger}_{\bm{k}_{\parallel},x_i+1}\bm{c}_{\bm{k}_{\parallel},x_{i}} + \bm{c}^{\dagger}_{\bm{k}_{\parallel},x_i}\bm{c}_{\bm{k}_{\parallel},x_{i}-1} \Big\}\bigg].
\end{align}
By following the linear response theory, the expectation value of the paramagnetic current is given by
\begin{eqnarray}
    \langle \hat{J}^{\rm para}_y(x_i)\rangle = - \frac{1}{c_0}\sum_{x_j}K^{\rm para}_y(x_i,x_j) A_y(x_j),
    \label{eq:jpara}
\end{eqnarray}
where the paramagnetic kernel is defined as
\begin{align}
    K_y^{\rm para}(x_i,x_j) =&  -\frac{1}{2}\sum_{\bm{k}_{\parallel}}\sum_{\alpha,\beta}\frac{n_{\rm F}(E_{\beta,{\bm{k}}_{\parallel}+\delta\bm{q}}) - n_{\rm F}(E_{\alpha,{\bm{k}_{\parallel}}})}{E_{\beta,{\bm{k}_{\parallel}+\delta\bm{q}}}-E_{\alpha,{\bm{k}_{\parallel}}} + i\delta_{\varepsilon}} \nn \\
    & \times\bra{u_{\beta,{\bm{k}_{\parallel}+\delta\bm{q}}}}\hat{J}_{y}^{\rm para}(x_i)\ket{u_{\alpha,{\bm{k}_{\parallel}}}}\nn \\
    & \times\bra{u_{\alpha,{\bm{k}_{\parallel}}}}\hat{J}_{y}^{\rm para}(x_j)\ket{u_{\beta,{\bm{k}_{\parallel}+\delta\bm{q}}}}.
    \label{eq:Kpara}
\end{align}
Here, $E_{\bm{k}_\parallel}^\alpha$ and $\ket{u_{\alpha,{\bm{k}_{\parallel}}}}$ are the eigenvalue and eigenvector of the lattice BdG Hamiltonian in Eq.~\eqref{eq:Hbdg_lattice}, $H_{\rm BdG}^{\rm lattice}(\bm{k}_{\parallel})$,
labeled by the band indices $\alpha,\beta$. 
The function $n_{\rm F}(E) = [e^{E/T}+1]^{-1}$ is the fermi distribution function at temperature $T$. In Eq.~\eqref{eq:Kpara}, we have also introduced an infinitesimal positive constant, $\delta_{\varepsilon}$.
The diamagnetic kernel is given by
\begin{align}
    K_y^{\rm dia}(x_i) = \frac{1}{2}\sum_{\bm{k}_{\parallel},\alpha}n_{\rm F}(E_{\alpha,{\bm{k}_{\parallel}}})
    \bra{u_{\alpha,{\bm{k}_{\parallel}}}}\hat{K}^{\rm dia}_y(x_i)\ket{u_{\alpha,{\bm{k}_{\parallel}}}}.
\end{align}

\subsection{Maxwell's equation}
In order to obtain the spatial distribution of the magnetic field inside the superconductor, we solve Maxwell's equation for the vector potential $A_y(q_x)$ by employing the Fourier transformation. However, we must be cautious because the Fourier-transformed response functions, $K^{\rm para}_y(q_x)$ and $K^{\rm dia}_y(q_x)$, are defined in an infinite space that includes the vacuum region.
To address this issue, we employ a method that introduces sheet-like source currents in an infinite medium filled with a periodic array of superconducting domains with the mirror reflected vector potential and magnetic field~\cite{tinkham1996superconductivity}. Figure~\ref{Fig:mirror}(b) shows the schematic image of our setup.  
We first consider the superconducting domain that occupies the region $x_i = 1\dots L$. Then, we introduce a periodic array of the mirror-reflected images of the superconducting domain. Each superconducting domain is separated by sheets through external current flows as $J_{y}^{\rm ex} = \pm(c_0/2\pi)B_0$, associated with the discontinuity of the magnetic field, $2B_0$. This external current sheet is crucial to ensuring the boundary conditions, whereby the internal field matches the applied external field at the surfaces of the superconducting domain.

Maxwell's equation for the periodic array of the superconducting domain and interface with a sheet current can be solved by employing the Fourier transformation. We start to discretize Maxwell's equation in this configuration as
\begin{align}
    &[A_y(x_{i}+1) - 2A_y(x_i) + A_y(x_{i}-1)]/a^2 \nn \\
    &= - \frac{4\pi}{c_0}\left[J_y^{\rm ex}(x_i) + J_y^{\rm dia}(x_i) + J_y^{\rm para}(x_i)\right].
\end{align}
The total current consists of three different contributions.
The external current associated with the applied magnetic field $B_0$ is defined by
\begin{align}
    J^{\rm ex}_y(x_i) = \frac{c_0}{4\pi}\sum_n [-2B_0\delta_{x_i,nN} + 2B_0\delta_{x_i,(n+1/2)N}],
\end{align}
where $N = 2L+2$ is the number of lattice sites in one supercell.
The paramagnetic current is obtained from Eq.~\eqref{eq:jpara} as
\begin{align}
    J^{\rm para}_y(x_i) = -\frac{1}{c_0}\sum_{x_j\in \text{SC}_{x_i}} K^{\rm para}_y(x_i,x_j)A_y(x_j).
\end{align}
Here, SC$_{x_i}$ denotes the set of sites within a superconducting domain.
The diamagnetic current is given by
\begin{align}
    J^{\rm dia}_y(x_i) = - \frac{1}{c_0}K^{\rm dia}_y(x_i)A_y(x_i).
\end{align}
By employing the Fourier transformation of the discretized Maxwell's equation, we obtain 
\begin{align}
    &\frac{2}{a^2}(\cos q_xa -1)A_y(q_x) = 2B_0(1-e^{-iq_xN/2}) \nn \\
    &+ \frac{4\pi}{c_0^2} \frac{1}{N}\sum_{q_x'}\Big[K^{\rm para}_y(q_x,-q_x') + K^{\rm dia}_y(q_x-q_x')  \Big]A_y(q_x'),
\end{align}
where we define Fourier-transformed response functions
\begin{gather}
    K^{\rm para}(q_x,q_x') = \sum_{x_i,x_j=1}^N e^{-iq_xx_i-iq_x'x_j}K^{\rm para}_y(x_i,x_j), \\
    K^{\rm dia}_y(q_x) = \sum_{x_i = 1}^{N}e^{-iq_xx_i}K^{\rm dia}_y(x_i).
\end{gather}
Here, the momentum $q_x$ takes discrete values as $q_x = 2\pi n/N$ with $n=1,\cdots,N$.
By solving the above equation, we obtain the vector potential $A_y(q_x)$, 
which can be used in Eq.~\eqref{eq:B} to reconstruct the spatial profile of the magnetic field $B_z(x_i)$.
The magnetic penetration depth $\lambda_y$ is then defined as the spatial integral of the magnetic field from the surface to the center of the slab:
\begin{eqnarray}
    \lambda_y(T) = \frac{1}{B_0} \int^{L/2}_0 dx~B_z(x).
\end{eqnarray}

We comment on the material parameters that influence the magnitude of the magnetic penetration depth.
In single-band superconductors, where the normal-state Hamiltonian consists only of Eq.~\eqref{eq:epsilon},
the paramagnetic response vanishes at zero temperature, and the diamagnetic response governs the magnetic field screening.
In this model, the London penetration depth $\lambda_{y}^{\rm London}$ can be defined in terms of the diamagnetic kernel as
\begin{align}
    [\lambda_{y}^{\rm London}]^{-2} &= \frac{4\pi}{c_0^2}K^{\rm dia}_y(q_x=0) \nn \\
    &= \frac{4\pi e^2 t_2b^2 n}{c_0^2\hbar^2},
\end{align}
where $n$ is the electron number density operator and
$t_2$ is the hopping integral introduced in Eq.~\eqref{eq:epsilon}.
Since the expression involves $t_2$, the magnitude of the magnetic penetration depth is related to the energy scale of the electronic system, with $t_2\sim W$.

\section{Effect of inter-orbital transitions on the penetration depth: Bulk without surface states}
\label{sec:bulk}

Before discussing the impact of Majorana surface states on the magnetic penetration depth, we begin by examining the behavior in the bulk without surface states.
We define the penetration depth in a spatially uniform superconducting system without surfaces as 
\begin{align}
    \lambda_{\rm Bulk}^{-2} = \frac{4\pi}{c_0^2}K_{\rm Bulk}(\bm{q}={\bm 0}).
\end{align}
The total Meissner kernel in bulk superconductors, $K_{\rm Bulk}(\bm{q}={\bm 0})$, is composed of paramagnetic and diamagnetic contributions:
\begin{align}
    K_{\rm Bulk}(\bm{q}={\bm 0}) = K^{\rm para}_{\rm Bulk} + K^{\rm dia}_{\rm Bulk}.
\end{align}
The paramagnetic part is given by
\begin{align}
    \label{eq:para_bulk}
    K^{\rm para}_{\rm Bulk} =& \frac{e^2}{2\hbar^2}\sum_{\bm{k},\alpha,\beta}
    \frac{n_{\rm F}(E_{\beta,{\bm{k}}}) - n_{\rm F}(E_{\alpha,{\bm{k}}})}{E_{\beta,{\bm{k}}}-E_{{\alpha},{\bm{k}}} + i\delta_{\epsilon}} \nn \\
    &\times \bra{u_{\beta,{\bm{k}}}}\partial_{k_y}\tilde{H}_{\rm BdG}(\bm{k})\ket{u_{\alpha,{\bm{k}}}} \nn \\ 
    &\times \bra{u_{\alpha,{\bm{k}}}}\partial_{k_y}\tilde{H}_{\rm BdG}(\bm{k})\ket{u_{\beta,{\bm{k}}}},
\end{align}
while the diamagnetic part is
\begin{align}
    K^{\rm dia}_{\rm Bulk} = \frac{e^2}{2\hbar^2}\sum_{\bm{k},\alpha}n_{\rm F}(E_{\alpha,{\bm{k}}})
    \bra{{u_{\alpha,{\bm{k}}}}}\partial^2_{k_y}\tilde{H}_{\rm{BdG}}(\bm{k})\ket{u_{\alpha,{\bm{k}}}}.
\end{align}

\begin{figure}[t]
    \centering
    \includegraphics[width=\linewidth]{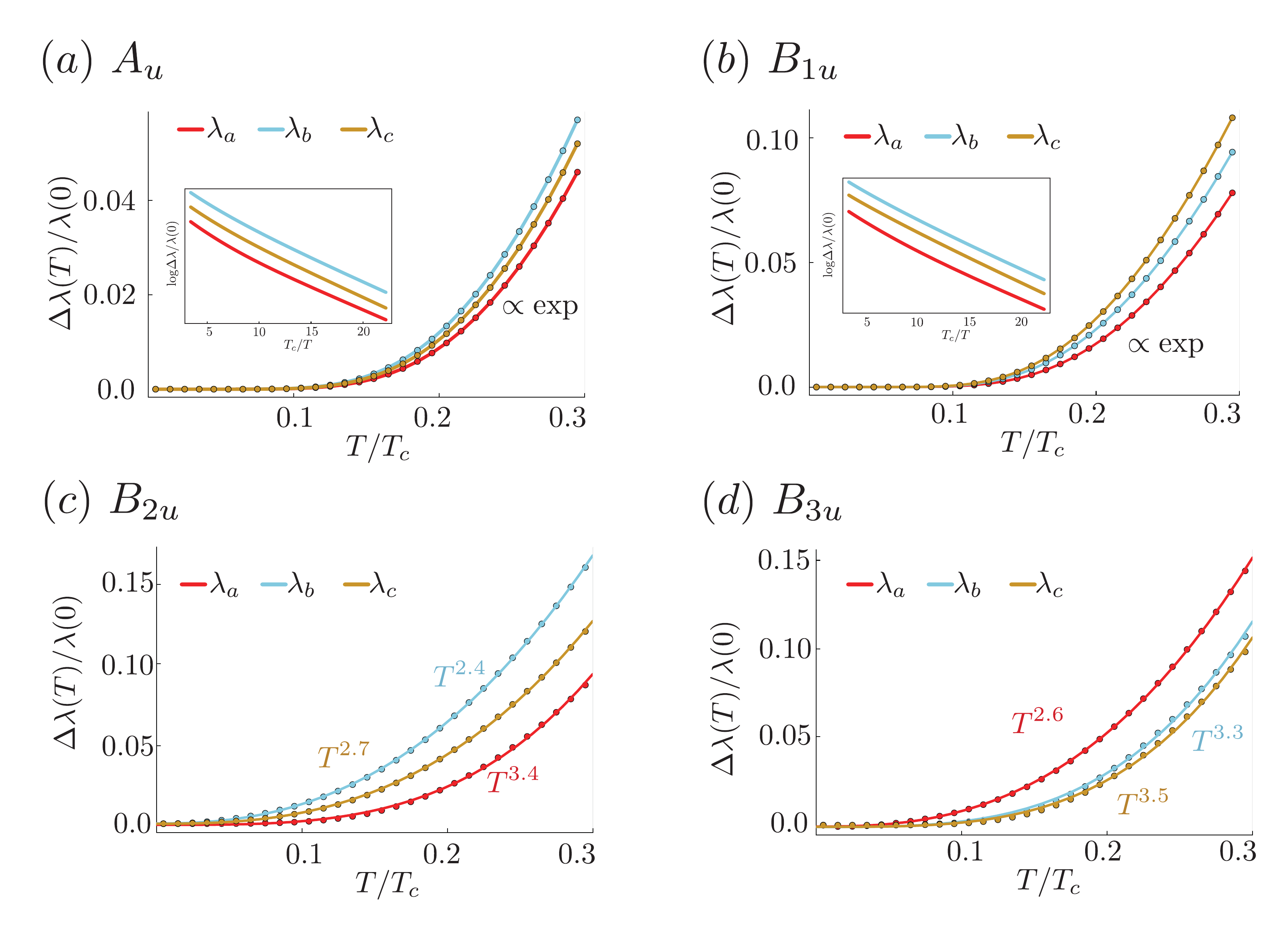} 
    \caption{Temperature dependence of magnetic penetration depth estimated from the bulk Meissner kernel.
    The insets in (a) and (b) show that $\log(\Delta\lambda)$ is proportional to the temperature, suggesting that exponential temperature behavior of $\Delta\lambda$.
    $n$ denotes the exponent in the power-law fitting $\Delta\lambda \propto T^n$.
    }
    \label{Fig:normal}
\end{figure}

Figure~\ref{Fig:normal} shows the temperature dependence of the magnetic penetration depth 
calculated from the bulk Meissner kernel for the $A_u$, $B_{1u}$, $B_{2u}$, and $B_{3u}$ states, respectively.
In the fully gapped $A_u$ and $B_{1u}$ states, the penetration depth decreases exponentially with decreasing temperature.
The semilogarithmic plot of the penetration depth appears in the insets of Figs.~\ref{Fig:normal}(a) and \ref{Fig:normal}(b), demonstrating that $\log\Delta(\lambda)$ is proportional to $1/T$. 

Let us now shift our attention to the penetration depth in the nodal $B_{2u}$ state.
Figure~\ref{Fig:normal}(c) shows the temperature dependence of $\Delta\lambda(T)=\lambda(T)-\lambda(0)$ for currents along each spatial direction, where $\lambda(0)$ is the magnetic penetration depth at zero temperatures. 
In all directions, $\Delta \lambda(T)$ exhibits a power-law behavior as $\Delta\lambda(T)\propto T^n$. We find that the exponents are
$n={3.4}$ for $\lambda_a$, $n={2.4}$ for $\lambda_b$, and $n={2.7}$ for $\lambda_c$, indicating the presence of gapless quasiparticle excitations at the nodal points.
According to conventional theory~\cite{einzel86,gross86}, the temperature dependence of the magnetic penetration depth in point nodal superconducting states is expected to be $T^2$ for the current $J_b$ along the nodal direction, and $T^4$ for $J_a$ and $J_c$.
The exponent for $\lambda_c$, which we present in Figs.~\ref{Fig:normal}(c), significantly deviates from this expectation.
It should be noted that the conventional theory is based on a model for single-band electrons with a spherical Fermi surface.
In such a model, the $B_{2u}$ state hosts point nodes at $k_a = k_c = 0$, and the low-energy quasiparticles carry momentum predominantly along the $b$-axis,
contributing only to $J_b$.
In contrast, the Hamiltonian for UTe$_2$, which is described in Sec.~\ref{sec:bdg}, has a cylindrical Fermi surface and also incorporates the orbital degrees of freedom of U atoms.
Below, we will demonstrate that these features can lead to differences in $\Delta \lambda (T)$ compared to the conventional predictions.

\begin{figure}[t]
    \centering
    \includegraphics[width=\linewidth]{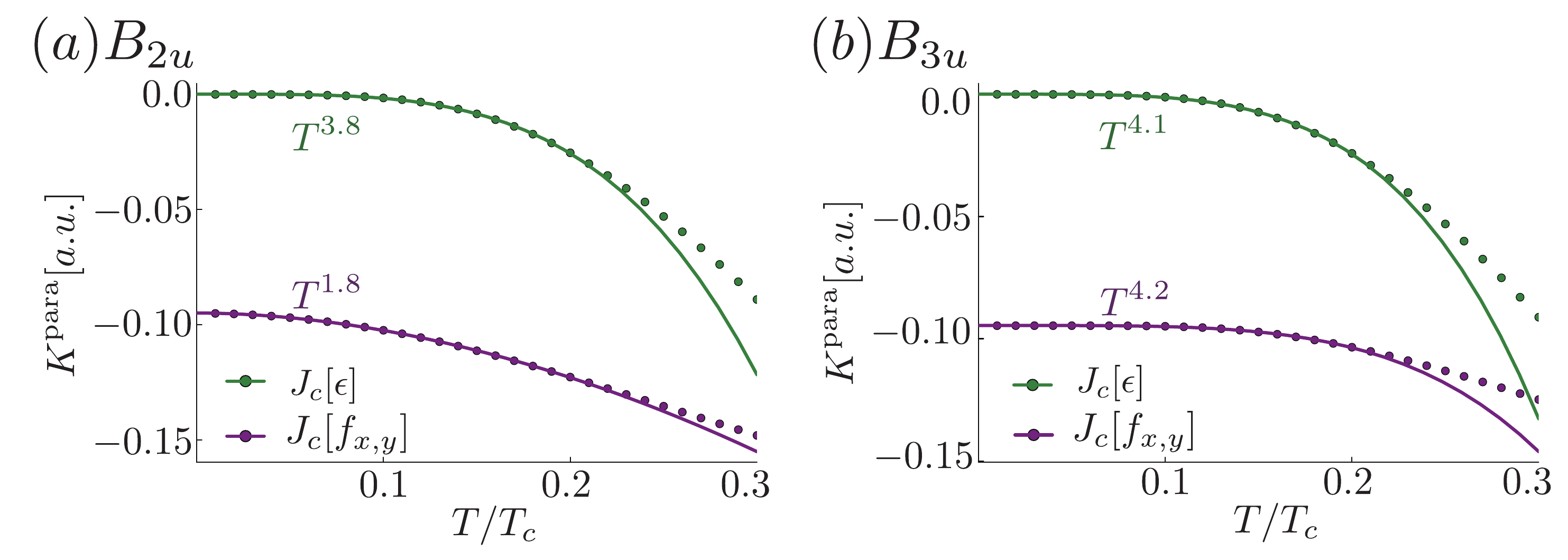} 
    \caption{Temperature dependences of paramagnetic Meissner kernel in the bulk $B_{2u}$ state (a) and $B_{3u}$ state (b), where $J_c[\epsilon]$ and $J_c[f_{x,y}]$ correspond to the intra-orbital and inter-orbital components of the paramagnetic Meissner kernel.
    $n$ denotes the exponent in the power-law fitting $\Delta\lambda \propto T^n$.}
    \label{Fig:Jinter}
\end{figure}

The temperature dependence of the magnetic penetration depth is governed by the paramagnetic response of thermally excited quasiparticles.
As shown in Eq.~\eqref{eq:para_bulk}, the paramagnetic current is proportional to $\partial_{k_c}H_N$, where the current is assumed to flow along the $c$-directions.
By substituting Eq.~\eqref{eq:Normal} into Eq.~\eqref{eq:para_bulk},
the paramagnetic current can be decomposed into two contributions:
the intra-orbital component $J_c[\epsilon_0(\bm{k})]$,
and the inter-orbital components $J_c[f_{x}(\bm{k})]$ and $J_c[f_{y}(\bm{k})]$.
Figure~\ref{Fig:Jinter}(a) shows the temperature dependences of the intra-orbital and inter-orbital components of the paramagnetic kernel for $J_c$ in the $B_{2u}$ state.
The intra-orbital kernel exhibits a power-law temperature dependence of $T^{3.8}$,
which is consistent with conventional expectations $T^4$.
In contrast, the inter-orbital kernel shows a lower exponent of $T^{1.8}$, resulting in a significant deviation from $\Delta \lambda(T)\propto T^4$.
We note that although the paramagnetic kernel also contains cross terms between $J_c[\epsilon_0(\bm{k})]$ and $J_c[f_{y}(\bm{k})]$, their contributions are small and negligible.

To reveal the origin of this diminished exponent in the inter-orbital paramagnetic response, let us now focus on the low-lying quasiparticle excitations near the nodal points at $k_a=0$ and $k_c=0, \pm \pi$, and discuss their contributions to the current.
The intra-orbital component $J_c[\epsilon_0] \propto \sin k_c$ vanishes at $k_c = 0$, implying that low-lying quasiparticles around the nodal points do not contribute to the paramagnetic response. This leads to the conventional power law, $T^4$, expected from a single-band model with a spherical Fermi surface.
On the other hand, low-lying quasiparticle excitations near the point nodes contribute a finite amount to the inter-orbital paramagnetic current, as 
\begin{align}
J_c[f_{y}(\bm{k})]\propto \cos(k_aa/2)\cos(k_bb/2)\cos(k_cc/2). 
\label{eq:jc}
\end{align}
Similarly, quasiparticle excitations in the vicinity of $k_a=0$ and $k_c=\pm\pi$ have nonvanishing contributions to the inter-orbital current $J_c[f_{x}(\bm{k})]$. 
The contribution of quasipaticles around the nodal points via the inter-orbital channel deviates the power-law behavior from $T^4$, leading to $\Delta \lambda (T)\propto T^2$. Furthermore, the Fermi surface of UTe$_2$ has a quasi-2D cylindrical shape opened to the $c$-axis, and the Fermi velocity in the $c$-axis is vanishingly small, which also suppresses the intra-orbital paramagnetic current along the $c$-axis ($J_c$). 
As a result, for $\lambda_c$ in the $B_{2u}$ state, the inter-orbital paramagnetic current carried by nodal quasiparticle becomes predominant, leading to a reduced exponent of $\Delta\lambda(T)$.

Lastly, we discuss the magnetic penetration depth in the $B_{3u}$ state. The nodal points are located at $k_b=\pm \pi$ on the electron Fermi surfaces, see Fig.~\ref{Fig:Fermi}(d).
As shown in Fig.~\ref{Fig:normal}(d), the numerical results are generally consistent with conventional predictions.
This implies that the contribution of the inter-orbital current is relatively small.
As shown in Fig.~\ref{Fig:Jinter}(b), the paramagnetic kernels for both intra-orbital and inter-orbital components of $J_c$ exhibit the conventional power-law, $T^4$, implying that low-energy quasiparticles do not contribute to the inter-orbital current.
This is because the nodes in the $B_{3u}$ state are located at $k_b = \pm \pi$,
where the inter-orbital current component in Eq.~\eqref{eq:jc} is proportional to $\cos k_bb/2$ and vanishes.
This suppression is entirely accidental and stems from the simplicity of the model.
In fact, UTe$_2$ possesses not only electron Fermi surfaces but also hole Fermi surfaces,
on which the inter-orbital current is expected to play a more significant role.

As shown by the above discussion, the power-law exponent of the temperature dependence of the magnetic penetration depth is highly sensitive to model details,
such as the shape of the Fermi surface and the presence of orbital degrees of freedom.
In the following subsections, the effect of Majorana modes is defined as the deviation from the bulk behavior discussed above.

\section{Effect of Majorana surface states on the penetration depth}
\label{sec:majorana}

\begin{figure*}[t]
    \centering
    \includegraphics[width=\linewidth]{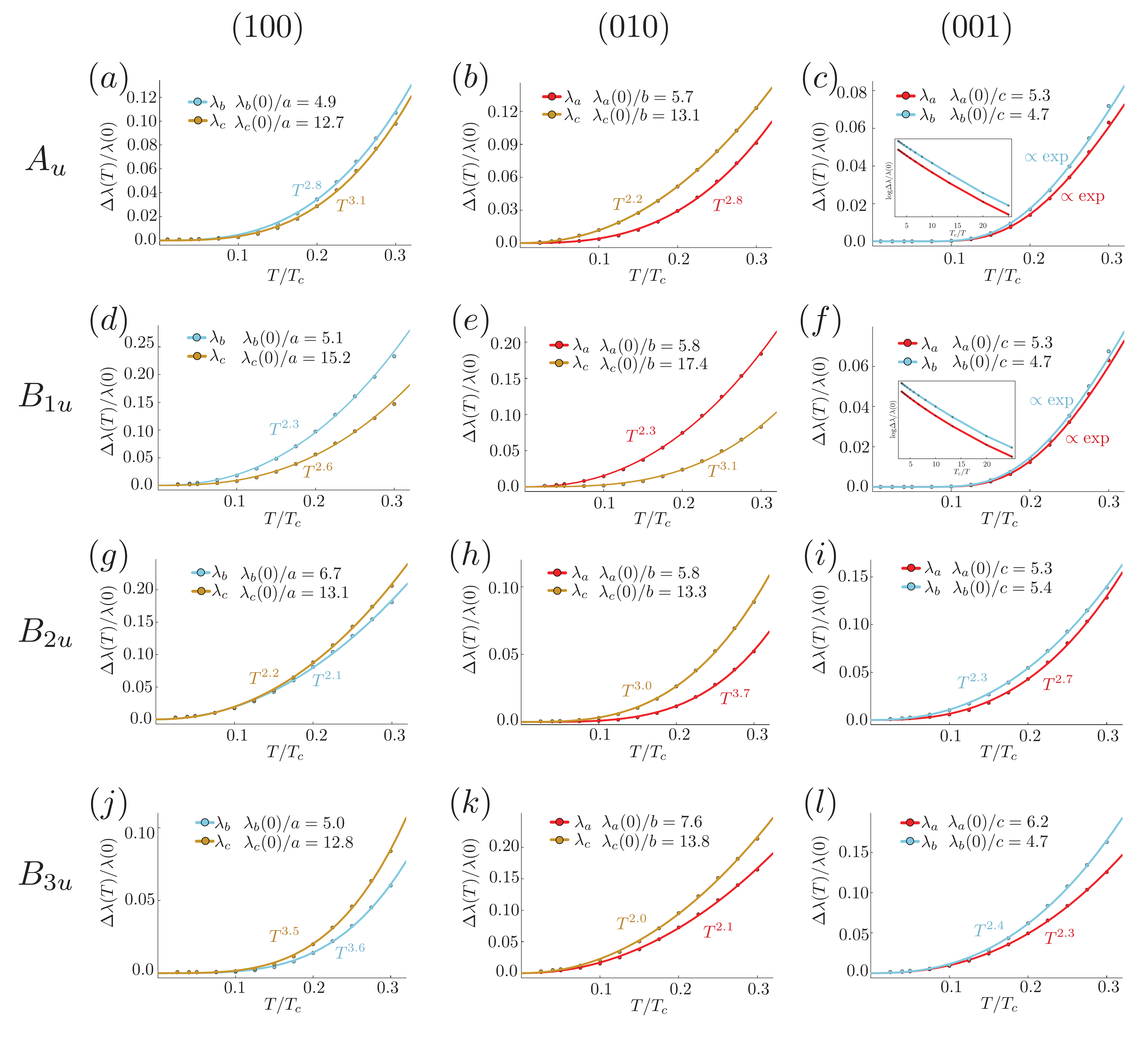} 
    \caption{Temperature dependences of the scaled magnetic penetration depth, $\Delta\lambda(T)/\lambda(0)$, for the $A_u$ state (a-c), the $B_{1u}$ state (d-f), the $B_{2u}$ state (g-i), and the $B_{3u}$ state (j-l).
    The first, second, and third columns show the $(100)$, $(010)$, and $(001)$ surfaces, respectively.
    Each curve is fitted either by a power-law or an exponential function.
    $n$ denotes the exponent in the power-law fitting $\Delta\lambda \propto T^n$.
    The insets in (c) and (f) confirm that $\Delta\lambda(T)$ is proportional to  $1/T$.}  % 全体のキャプション
    \label{Fig:all}
\end{figure*}

In this section, we present the results of numerical calculations and analysis of magnetic penetration depths for slab geometries cleaved along crystallographic planes, $(100)$, $(010)$, and $(001)$, for each IR [See Fig.~\ref{Fig:mirror}]. Figure~\ref{Fig:all} shows how the change in the magnetic penetration depth, $\Delta\lambda(T)=\lambda(T)-\lambda(0)$, from its zero-temperature value varies with temperatures for all IRs and the surfaces considered.

To maximize the effect of Majorana surface states, we first focus on the small $\kappa$ regime, and the type-II limit will be discussed in Sec.~\ref{sec:typeII}. 
In Secs.~\ref{sec:au} - \ref{sec:b3u}, the parameter $W$ was chosen such that both $\lambda_a$ (normalized by $b$ or $c$) and $\lambda_b$ (normalized by $a$ or $c$) are approximately six at $T=0$, comparable to or greater than the coherence length.
These parameters correspond to the low $\kappa$ regime, where $\kappa$ is the GL parameter defined by $\kappa_{i} \equiv \lambda_{i}(0)/\xi_i $ ($i=a,b,c$). 
Due to the anisotropy of the Fermi surface---particularly the small dispersion along the $c$-axis---the magnetic penetration depth is significantly longer in the $c$-axis direction. As shown in Fig.~\ref{Fig:all}, the ratio is $\lambda_c/\lambda_{a,b} \sim 2$ -- $3$. 

In the following subsections, we present the results of $\Delta\lambda(T)$ for each IR in detail. Then, we discuss how Majorana surface states affect the magnetic penetration depth in type-II superconductors, where the penetration depth is longer than the coherence length.

\subsection{{Impact of Majorana arcs endpoints on power laws}}
\label{sec:arcs}

\begin{figure}[t]
    \centering
    \includegraphics[width=0.9\linewidth]{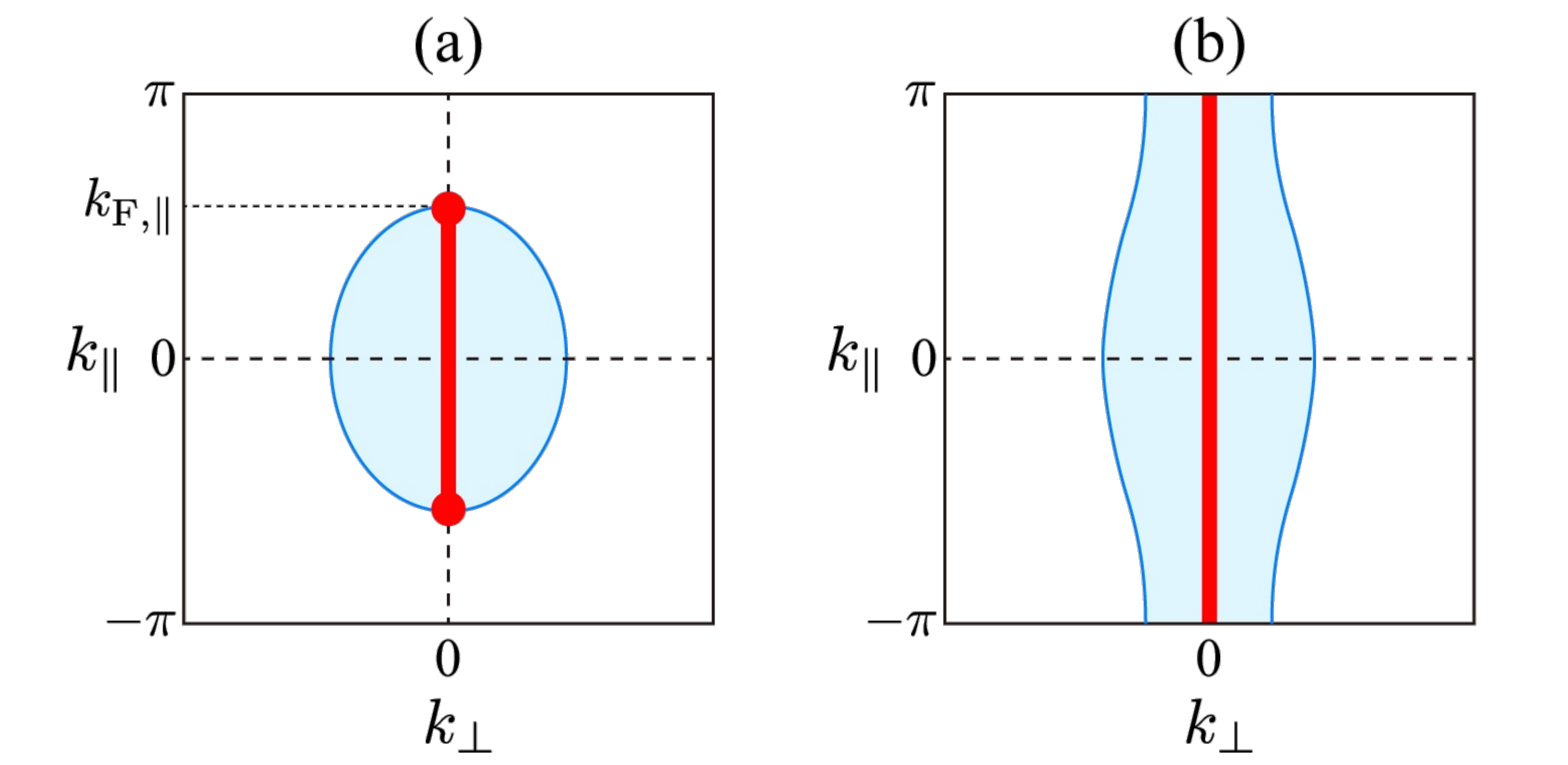} 
    \caption{{Schematics of Majorana arcs (a) with and (b) without endpoints in the surface Brillouin zone.}}  
    \label{Fig:arc}
\end{figure}

{Before presenting the results for each IR, we now examine how the power-law behavior of $\Delta\lambda\propto T^n$ associated with Majorana arcs depends on the direction of the screening current and on the presence or absence of arc endpoints. We will show that the difference of the power law exponent in $B_{1u}$, $B_{2u}$, and $B_{3u}$ states are qualitatively well understandable with the flow direction and the presence or absence of Majorana arc endpoints in the surface Brillouin zone. Technical details of the derivation are supplemented in Appendix~\ref{appendix}.}

{As shown in Fig.~\ref{Fig:arc}, we consider two typical situations of Majorana arcs: those with and without endpoints. The former (latter) corresponds to the surface Majorana arcs in $B_{2u}$ and $B_{3u}$ ($B_{1u}$) states. We set $k_{\parallel}$ and $k_{\perp}$ to be the momenta parallel and perpendicular to the direction of the Majorana arc, respectively.}

{The presence or absence of the Majorana arc endpoints has a direct impact on the temperature dependence of the magnetic penetration depth along the arc. As demonstrated in Ref.~\cite{wu21}, when the Majorana arc terminates at well-defined endpoints, the phase-space restriction near the endpoints in the current-current correlation function leads to a robust $T^2$ dependence of the surface contribution, regardless of the direction and the length of the Majorana arc. Thus, the penetration depths due to the screening current flowing along $k_{\perp}$ and $k_{\parallel}$ directions exhibit the following power laws
\begin{align}
\Delta \lambda _{\perp} \sim T^2, \quad
\Delta \lambda _{\parallel} \sim T^2.
\end{align}
However, as we discuss in Appendix~\ref{appendix}, when the endpoints are absent and the arcs extend across the surface Brillouin zone, the exponent remains $n=2$ in the $k_{\perp}$ direction, while deviating from $n=2$ in the $k_{\parallel}$ direction. Therefore, the temperature dependence of the penetration depth due to Majorana arcs without endpoints read
\begin{gather}
\Delta \lambda _{\perp} \sim T^2, \\
\Delta \lambda _{\parallel} \sim a T^2 + be^{-\tilde{E}_{\rm min}/T}.
\end{gather}
The characteristic energy scale is expressed by $\tilde{E}_{\rm min}=O(0.1\Delta)$.
}

{In summary, the temperature dependence of the magnetic penetration depth in superconductors hosting surface Majorana arcs is given by 
\begin{align}
\Delta \lambda_{\perp} \sim T^2, \quad \Delta \lambda_{\parallel} \sim T^{2+\eta(T)},
\end{align}
where $\eta>0$ is not universal but depends on the material details and $T$. We note that $\Delta \lambda _{\perp} \sim T^2$ follows solely from the linear surface dispersion in the $k_{\perp}$ direction, and therefore remains unchanged even if the detailed shape of the Majorana arc is altered--for instance, by removing the arc endpoints and by changing the arc length. For $\Delta \lambda_{\parallel}$, the additional exponent vanishes ($\eta=0$) when the arc has endpoints [e.g., the (100) surface in the $B_{2u}$ state and the (010) surface in the $B_{3u}$ state]. When the arc endpoints are absent [e.g., the (100) and (010) surfaces in the $B_{1u}$ state], the leading temperature dependence is controlled by the robust $T^2$ contribution, with a nonuniversal crossover near $T\sim E_{\min}$ due to the thermally activated contribution.}

{Using the parameters of our tight-binding model defined in Sec.~\ref{sec:bdg}, the characteristic temperature associated with this crossover can be estimated as $T^\ast \sim E_{\min} \approx 0.3T_c$. On the other hand, as discussed in the following subsections, the power of $T$ in the numerical results is extracted by fitting the data in the temperature range $T/T_c \sim 0.1$–$0.3$. This interval, therefore, lies close to the crossover scale. As a result, for $\Delta \lambda_c$ in $B_{1u}$, the apparent temperature dependence obtained from numerical calculations deviates from the asymptotic $T^2$ behavior, reflecting the gradual onset of the thermally activated contribution discussed above.
}

\subsection{The \texorpdfstring{$A_u$}{Au} pairing state}
\label{sec:au}
The $A_u$ state hosts Majorana surface states with a cone-shaped dispersion on the $(100)$ and $(010)$ surfaces, while no surface states appear on the $(001)$ surface.
Figures~\ref{Fig:all}(a-c) show $\Delta\lambda(T)$ for each surface orientation.
For the $(001)$ surface, $\Delta\lambda_a(T)$ and $\Delta\lambda_b(T)$ exhibit an exponential decay with temperature, consistent with the bulk behavior in Sec.~\ref{sec:bulk}.

In contrast, the $(100)$ and $(010)$ surfaces show power-law temperature dependences in the penetration depth, $\Delta \lambda (T)\propto T^n$. The estimated exponents of $\Delta\lambda_{b,c}$ on the $(100)$ surface and for $\Delta \lambda_a$ for the $(010)$ surface are approximately $n\approx 3$ or slightly less, in good agreement with prior analytical results obtained from the continuum model for a fully gapped topological superconducting state with a Majorana cone~\cite{wu2020power}.
Note that the exponent for the $\lambda_c$ on the $(010)$ surface is found to be close to $n=2$ [see Fig.~\ref{Fig:all}(b)].
As discussed in the previous subsection, the intra-orbital component of the paramagnetic current response is strongly suppressed because of the quasi-2D structure of the Fermi surface. On the other hand, the inter-orbital paramagnetic current becomes predominant in $J_c$, leading to a decrease in the power-law exponent, $n<3$.

\subsection{The \texorpdfstring{$B_{1u}$}{B1u} pairing state}

The $(001)$ surface in the $B_{1u}$ state is topologically trivial and not accompanied by Majorana surface states. The magnetic penetration depths for this surface orientation exhibit an exponential temperature dependence,
as shown in Fig.~\ref{Fig:all}(f).
A key difference from the $A_u$ state is that the $B_{1u}$ state hosts the Majorana arcs on both the (100) and (010) surfaces, which are dispersive in one direction and form a zero-energy flatband in the other. In contrast to the cone-shaped dispersion in the $A_u$ state, these Fermi arcs result in a large amount of surface density of states at zero energy. The quantitative difference in the density of states of the surface Majorana modes causes a change in the power-law exponent of the magnetic penetration depth, shifting from the exponent $n=3$ for an isotropic Majorana cone. 

Figure~\ref{Fig:all}(d) shows the temperature dependences of $\Delta \lambda_b(T)$ and $\Delta\lambda_c(T)$ on the $(100)$ surface.
The penetration depth, $\Delta \lambda(T)$, exhibits a temperature dependence following a power law such as $\Delta\lambda_b \propto T^{2.3}$ and $\Delta \lambda_c(T)\propto T^{2.6}$.
As noted in Table~\ref{table:topo}, the Majorana band that forms on the $(100)$ surface is dispersionless along the $k_c$-direction, which is protected by the mirror reflection symmetry in the $ca$ plane, $M_{ca}$. The Majorana surface state contributes less to the paramagnetic current in $J_c$ than to $J_b$. As a result, the dispersionless surface states along the $c$-axis exhibit a reduced paramagnetic contribution to the screening current, $J_c$, resulting in a larger exponent in $\Delta \lambda_c$ compared to $\Delta\lambda_b$.
A similar feature is also observed in the magnetic responses on the $(010)$ surface, where the Majorana band is dispersive along $k_a$ and flat along $k_c$.
For this surface orientation, as shown in Fig.~\ref{Fig:all}(e), $\Delta \lambda(T)$ exhibits a power-law behavior with $\Delta\lambda_a \propto T^{2.3}$ and $\Delta\lambda_c\propto T^{3.1}$.
Since the $B_{1u}$ state is fully gapped in the bulk, the low-temperature behavior is dominated by contributions from the Majorana Fermi arcs. {It is important to note that the Majorana arc has no endpoints and extends across the surface Brillouin zone. Therefore, as discussed in Sec.~\ref{sec:arcs}, these results indicate that in systems with Majorana Fermi arcs, the magnetic penetration depth tends to follow a power law with an exponent of approximately $n=2$ along the dispersive direction. In contrast, the exponent exceeds $n=2$ in the dispersionless direction. The anisotropic magnetic response directly reflects the protection of Majorana arcs in the absence of endpoints.}

\subsection{The \texorpdfstring{$B_{2u}$}{B2u} pairing state}
The $B_{2u}$ state hosts Majorana Fermi arc on the $(100)$ and $(001)$ surfaces, while no surface states appear on the $(010)$ surface.
Figures~\ref{Fig:all}(g-i) show the temperature dependence of $\Delta\lambda(T)$ for each of these surfaces.
The $(010)$ surface, where Majorana surface states are absent, exhibits a temperature dependence consistent with the bulk behavior as discussed in Fig.~\ref{Fig:normal}(c).
For the $(100)$ surface, the Majorana surface state is dispersive along the $k_c$-direction and flat along the $k_b$-direction. {We note that the Majorana arcs appear along $k_{\rm c}=0$ and $k_c=\pm \pi$ and are terminated at finite $k_{b}$. Therefore, as mentioned in Sec.~\ref{sec:arcs}, the existence of the arc endpoints ensures the robust $T^2$ dependence of the magnetic penetration depth.}
Indeed, the power-law exponent of $\lambda_c$ for the current along the dispersive direction, $J_c$, tends toward $n=2$, reflecting the influence of dispersive surface states on the paramagnetic current.
Although $J_b$ flows along the dispersionless direction of the Majorana arc, the exponent of $\Delta\lambda_b$ also remains around $n=2$, which is in contrast to $\Delta\lambda_c$ for the (100) and (010) surfaces in the fully gapped $B_{1u}$ state with the Majorana arc. {As quasiparticles near the bulk nodal poins give rise to anisotropic exponents, the isotropic $T^2$ behavior is attributed to the existence of the terminated Majorana arcs.} 

As mentioned in Table~\ref{table:topo}, the $(001)$ surface hosts the Majorana Fermi arc that is dispersive along the $k_a$-direction and flat along the $k_b$-direction.
As shown in Fig.~\ref{Fig:all}(i), although the current $J_a$ flows along the dispersive direction of the Majorana arc state, $\Delta\lambda_a$ exhibits the temperature dependence following $T^{2.7}$. As mentioned above, the screening current along the dispersive direction of the Majorana arc commonly exhibits the exponent $n\approx 2$. However, the exponent for the (001) surface in the $B_{2u}$ state is larger than the expected value, $n=2$. 
This enhancement is likely due to additional contributions from bulk quasiparticle excitations near the point nodes, $\Delta \lambda^{\rm bulk}_a \sim T^4$, in addition to those from the Majorana surface states.

\subsection{The \texorpdfstring{$B_{3u}$}{B3u} pairing state}
\label{sec:b3u}
The $B_{3u}$ state hosts Majorana Fermi arcs on the $(010)$ surface, while no Majorana surface states appear on the other surface orientations. {We note that $B_{2u}$ and $B_{3u}$ are related by exchanging the $a$ and $b$ axes and therefore lead to similar physical behavior up to a relabeling of crystallographic directions, except for the (001) surface.}
Figures~\ref{Fig:all}(j-l) show the temperature dependence of $\Delta\lambda(T)$ for each of these surfaces.
As the $(100)$ surface does not host Majorana states, its penetration depth exhibits a temperature dependence similar to the bulk behavior in Fig.~\ref{Fig:normal}(d).
The Majorana Fermi arc formed on the $(010)$ surface is dispersive along the $k_c$-direction but remains dispersionless along the $k_a$-direction.
{Similarly with the results in the (100) surface of the $B_{2u}$ state, the temperature dependence of the magnetic response exhibits $\Delta\lambda_a\propto T^{2.1}$ and $\Delta\lambda_c\propto T^{2.0}$ in the dispersive and dispersionless directions, respectively. These are consistent with the power-law exponent expected from the contributions of the Majorana arcs with endpoints.}

Special attention should be paid to the result for the $(001)$ surface.
Despite the absence of Majorana surface states on this surface, the power-law exponent for $\Delta\lambda_b$, $n=2.4$, is found to be anomalously lower than that in the bulk, $n=3.3$ [see Fig.~\ref{Fig:all}(l)].
We note that the (001) surface does not host zero-energy Majorana surface states, but it is accompanied by in-gap states, which are trivial Andreev bound states~\cite{ohashi2024anisotropic}.
The necessary condition for nontrivial topology is the odd-parity pairing function,
$\Delta(\bm{k})=-\Delta(\underline{\bm{k}})$ with $\underline{\bm k}\equiv {\bm k}-2{\bm k}(\hat{\bm k}\cdot\hat{\bm n})$, where we set $\hat{\bm k}={\bm k}/|{\bm k}|$ and the unit vector normal to the surface, $\hat{\bm n}$. In the $B_{3u}$ state, the gap function $d_b \propto \sin k_c$ satisfies this condition for the (001) surface.
However, the topological invariant is trivial and zero-energy Majorana surface states do not exist because the cylindrical Fermi surface does not intersect the mirror-invariant plane on which the invariant is defined.
Nonetheless, once a Fermi surface appears on a mirror-symmetric plane, Majorana zero-energy states appear, and it seems that a remnant of them appears in this calculation.
Indeed, Ref.~\cite{ohashi2024anisotropic} has demonstrated that the (001) surface in the $B_{3u}$ pairing state hosts trivial surface bound states that are absent in the bulk, and its surface density of states has a large amount of ingap states. These ingap states contribute to the anomalously lower exponent, $\Delta\lambda_b\propto T^{2.4}$.

\subsection{Type-II limit}
\label{sec:typeII}
\begin{figure}[t]
    \centering
    \includegraphics[width=\linewidth]{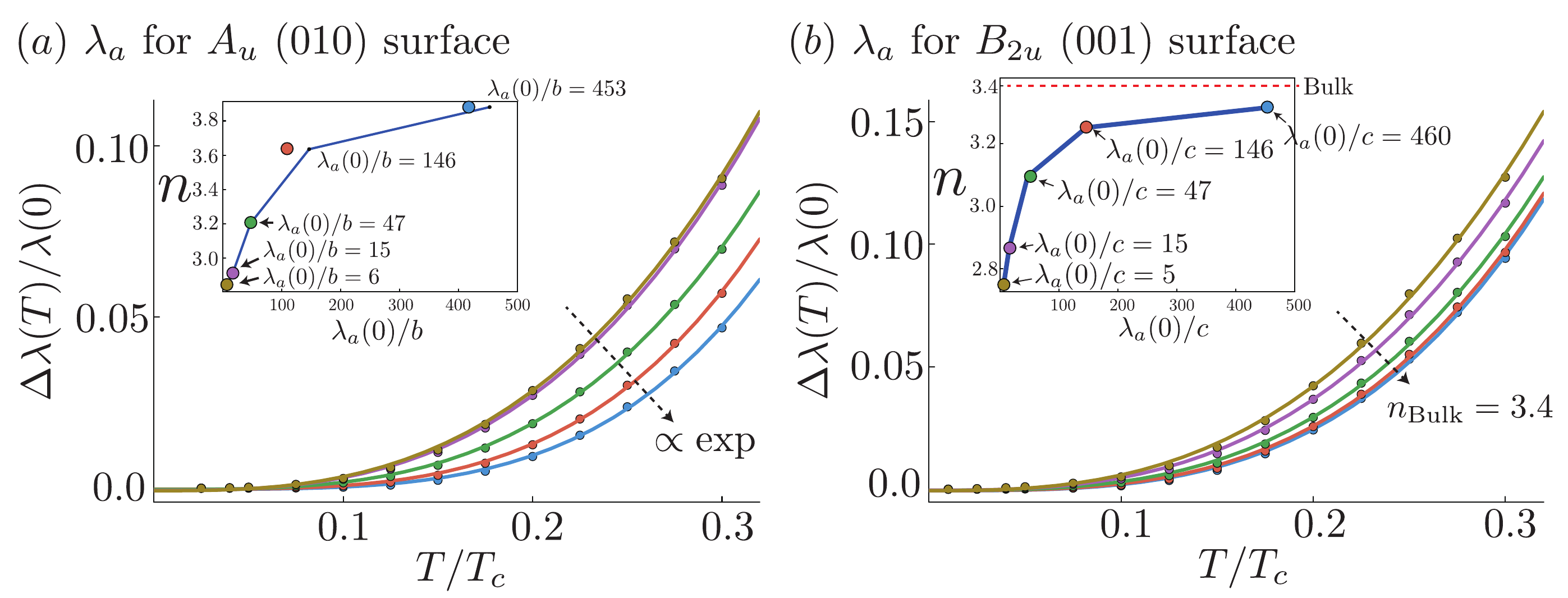} 
    \caption{(a)~Temperature dependence of $\Delta\lambda_a(T)$ for the $A_u$ state with $(010)$ surface.
    (b)~Temperature dependence of $\Delta\lambda_a(T)$ for the $B_{2u}$ state with $(001)$ surface.
    The insets show how the power-law exponent of the temperature dependence of $\Delta\lambda(T)$ varies with the zero-temperature value $\lambda(0)$.
    In the limit where the magnetic penetration depth $\lambda$ is much longer than the coherence length $\xi$, the temperature dependence of $\lambda$ approaches that of the bulk behavior.}  % 全体のキャプション
    \label{Fig:lambdachange}
\end{figure}

From the preceding discussions, we have demonstrated that Majorana surface states cause a significant change in the temperature dependence of the magnetic penetration depth when the magnetic penetration depth is comparable to the coherence length $\xi$.
However, UTe$_2$ is a type-II superconductor where $\lambda(0)$ is much longer than $\xi$ in all field directions, with $\lambda(0)$ in UTe$_2$ around $1~\mu{\rm m}$ and approximately $\xi =O(10~{\rm nm})$~\cite{ish23,bae21,met19} . Let us now examine the effect of Majorana surface states in the type-II limit, $\xi \ll\lambda(0)$.

Figure~\ref{Fig:lambdachange}(a) plots the temperature dependence of $\Delta\lambda_a(T)$ for the $(010)$ surface in the $A_u$ pairing state, for several values of $\lambda_a(0)/b$ ranging from {$6$} to $453$ {with a fixed $\xi_b/b=4$}. 
The inset shows how $\lambda_a(0)$ influences the power-law exponent, $n$, derived from the fitting of $\Delta \lambda_a(T)\propto T^{n}$.
In the low $\kappa$ regime where $\lambda(0)$ is comparable to $\xi$, the paramagnetic current due to the Majorana surface states dominates, leading to $T^3$ temperature dependence.
However, as $\kappa\equiv \lambda_a(0)/\xi_b$ increases, the contributions of the surface states to the paramagnetic current weaken and the diamagnetic current relatively strengthens. As a result, the exponent of $\Delta\lambda_a(T)$ gradually intensifies and shifts towards exponential behavior, similar to that of the bulk $A_u$ state.
The similar behaviors are observed in $\Delta \lambda_a(T)$ for the $(001)$ surface in the $B_{2u}$ state, which hosts the Majorana Fermi arc, as shown in Table~\ref{table:topo}.
It is seen from Fig.~\ref{Fig:lambdachange}(b) that as $\lambda_a(0)/\xi_c$ increases, the exponent of $\Delta\lambda_a(T)$ approaches the value in the bulk point-nodal pairing state, $n=3.4$. These results indicate that the contribution of Majorana surface states to the paramagnetic current becomes significant and significantly deviates from the temperature dependence of the magnetic penetration depth as $\lambda(0)$ approaches the coherence length $\xi$.

\section{Concluding remarks}
\label{sec:conclusion}

We have examined the effect of inter-orbital transitions and the impact of Majorana surface states on the temperature dependence of the magnetic penetration depths in topological superconductors. 
Here, we focus on UTe$_2$ as a specific example of the topological superconductor. We start with an effective model that incorporates the orbital degrees of freedom of UTe$_2$ and consider the superconducting states classified in terms of four irreducible representations of the $D_{2h}$ point group symmetry of the crystal: the $A_{u}$, $B_{1u}$, $B_{2u}$, and $B_{3u}$ pairing states. In UTe$_2$ with cylindrical Fermi surfaces, the $(100)$ and $(010)$ surfaces of the $A_{u}$ state host Majorana cones, while Majorana Fermi arcs protected by magnetic mirror symmetries appear on the particular surface orientations in the other pairing states.

First, in Sec.~\ref{sec:bulk}, we have discussed the inter-orbital effect of electrons on the magnetic penetration depth. In bulk nodal superconductors, such as the $B_{2u}$ and $B_{3u}$ states, the penetration depth along the antinodal direction exhibits a $T^2$ dependence, which significantly deviates from $T^4$ expected from conventional theory. This anomalous exponent is attributed to quasiparticle contributions around the point nodes to the inter-orbital paramagnetic current.

Furthermore, in Sec.~\ref{sec:majorana}, we have demonstrated that when the penetration depth at $T=0$, $\lambda(0)$, is comparable to the superconducting coherence length $\xi$, Majorana surface states significantly contribute to the paramagnetic current. As a result, the temperature dependence of the magnetic penetration depth deviates from that of bulk superconducting states. {In a simple single-band system with an isotropic Fermi sphere, the surface contribution can be qualitatively understood from the dimensionality of the Majorana spectrum: a Majorana cone typically gives an $n=3$ power law, whereas a Majorana arc leads to $n=2$ in any current directions~\cite{wu2020power,wu21}. We have extended the arguments in Ref.~\cite{wu21} to the case where the Majorana arcs have no endpoints. The exponent in the dispersive direction is robust against the change of the arc length and the presence or absence of the arc endpoints. However, we have demonstrated that when the Majorana arc is terminated at momenta in the surface Brillouin zone, the exponent in the dispersionless direction deviates from $n=2$. The presence or absence of the arc endpoints significantly alters the exponent of the screening current along the dispersionless direction at moderate temperatures.}

Then, we have discussed the exponents of the penetration depth based on the full numerical calcualtions of the effective tight-binding model for UTe$_2$. In the $A_u$ pairing state, the Majorana surface cone causes the $T^3$ power-law behavior of the penetration depth for both the (100) and (010) surfaces. 
{The (100) and (010) surfaces in $B_{1u}$ state show anisotropic magnetic responses. This is attributed to the existence of the Majorana arcs which are not terminated and extends across the surface Brillouin zone. For the (100) surface in the $B_{2u}$ state and the (010) surface in the $B_{3u}$ state, the existence of terminated Majorana arcs ensures the robust $n=2$ exponent in any current directions. All these behaviors in the $B_{1u}$, $B_{2u}$, and $B_{3u}$ states are consistent with the results obtained from the power-counting arguments in a single-band system.}
Therefore, in low-$\kappa$ type-II superconductors, the signals of Majorana surface states may be detectable through penetration depth measurements. As $\kappa$ increases and approaches the type-II limit, however, the effect of Majorana surface states on the paramagnetic current becomes weaker, and the temperature dependence of the magnetic penetration depth is governed by quasiparticle excitations in the bulk pairing states instead of surface states. In this limit, the quasiparticles around nodal points and the inter-orbital effect play a key role in determining the electromagnetic response. 

{We have not observed the non-monotonic temperature dependences of the magnetic penetration depth, unlike the case of $d$-wave superconductors~\cite{higashitani,walter,asano11,barash} and small unconventional superconductors~\cite{suz14,suz15}. In $d$-wave superconductors, due to the nodal line, the zero-energy flat band appears as a surface in the two-dimensional surface Brillouin zone, resulting in a prominent peak in the zero-energy density of states. In general, the zero-energy states exhibit negative superfluid density and give rise to the paramagnetic Meissner effect~\cite{tan05,yok11,tanaka12,hig14,lin19}. The large amount of zero-energy density of states in $d$-wave superconductors leads to an increase in the magnetic penetration depth with decreasing temperatures. In UTe$_2$, however, the possible superconducting states are either fully gapped or have point nodes, and the zero-energy states appear at a point or along a line within the surface Brillouin zone~\cite{tei2023possible,ohashi2024anisotropic} (see also Fig.~\ref{Fig:Majo}). As a result, the relatively small amount of zero-energy density of states in UTe$_2$ is insufficient to produce non-monotonic temperature dependences, leading instead to monotonic power-law behaviors as shown in this paper.}

The recent experiment in UTe$_2$~\cite{ish23} reveals that the magnetic penetration depth exhibits a $T^2$ dependence in all field directions. Although topological surface states play a significant role in the electromagnetic response of topological superconductors, they cannot solely explain the anomalous power-law behavior of the magnetic penetration depth observed in the superconductor UTe$_2$ with $\lambda(0)/\xi \gg 1$. Apart from the scenario of nonunitary pairing proposed in Ref.~\cite{ish23}, a key factor in understanding the anomalous electromagnetic response is the inter-orbital effect of electrons. As discussed in Sec.~\ref{sec:bulk}, for cylindrical Fermi surfaces, the diamagnetic current flowing along the cylinder is suppressed, whereas the paramagnetic current can be enhanced by nodal quasiparticles via the inter-orbital process. As a result, the inter-orbital paramagnetic current deviates the power-law behavior from $\Delta\lambda\propto T^4$ to $T^2$. This inter-orbital effect is highly sensitive to the electron band structure in the normal state, and further study based on a more sophisticated and realistic model remains as future work.

\section*{ACKNOWLEDGMENTS}
The authors are grateful to K. Ishihara and T. Shibauchi for valuable discussions from an experimental point of view.
K.A. and J.T. are grateful to Y. Yamazaki for fruitful discussions.
J.T. is supported by a Japan Society for the Promotion of Science (JSPS) Fellowship for Young Scientists.
This work was supported by JSPS KAKENHI (Grant No.~JP23K20828, No.~JP23K22492, No.~JP24KJ1621, No.~JP25H00599, No.~JP25H00609, No.~JP25K07227, No.~JP25K22011, and No.~JP25K23361) 
and a Grant-in-Aid for Transformative Research Areas (A) ``Correlation Design Science'' (Grant No.~JP25H01250) from JSPS of Japan.

\appendix

\section{{Paramagnetic kernel and Majorana arcs}}
\label{appendix}

{This Appendix provides a complementary analysis of the power-law behavior of the magnetic penetration depth discussed in Sec.~\ref{sec:arcs}, based on the paramagnetic kernel and power-counting considerations. Let us start by considering the current-current response function $K_{\mu\nu}$. We divide it into contributions from continuum states and Majorana arc states and focus on the latter. Then, the paramagnetic kernels in Eq.~\eqref{eq:Kpara}, $K_{\mu\nu,ij}^{\rm para}\equiv K_{\mu\nu}^{\rm para}(x_i,x_j)$, are given by~\cite{wu21} 
\begin{align}
    K_{\mu\nu,ij}^{\rm para}\propto&  \int \frac{d^2{\bm k}_{\rm s}}{(2\pi)^2} 
    J_{\mu}({\bm k}_{\rm s})J_{\nu}({\bm k}_{\rm s})
    \frac{\partial f(E_{{\bm k}_{\rm s}})}{\partial E_{{\bm k}_{\rm s}}}
    \Sigma^{\rm ss}_{ij}({\bm k}_{\rm s}),
    \label{eq:Kpara2}
\end{align}
where ${\bm k}_{\rm s}\equiv (k_{\perp},k_{\parallel})$ and $\Sigma^{\rm ss}_{ij}({\bm k}_{\rm s})=|\psi_{{\bm k}_{\rm s}}(x_i)|^2|\psi_{{\bm k}_{\rm s}}(x_j)|^2$. For simplicity, we neglect the orbital degrees of freedom and consider single-band systems with normal-electron dispersion, 
\begin{align}
\epsilon_0(x,{\bm k}_{\rm s})
=\frac{1}{2m_0}\left[-\partial^2_x + \alpha k^2_{\perp} 
+ \beta \cos(k_{\parallel})\right] - \epsilon_{\rm F}. 
\end{align}
The paramagnetic current operator then reduces to $\hat{J}_{\mu}=\partial_{k_{\mu}}\epsilon_0({\bm k})$. The Majorana arc with the dispersion, 
\begin{align}
E_{{\bm k}_s}= \frac{\Delta}{k_{\rm F}} k_{\perp}, 
\end{align}
appears within the momentum region $\alpha k^2_{\perp} + \beta\cos(k_{\parallel})\le k^2_{\rm F}$, and the zero energy states are dispersionless along the $k_{\parallel}$ diretion. The wavefunction reads 
\begin{align}
\psi_{{\bm k}_{\rm s}}(x)=N_{{\bm k}_{\rm s}}e^{-x_i/l_{\rm coh}}\sinh(\kappa x), 
\end{align} 
where we set $k^2_{\rm F}\equiv 2m_0\varepsilon_{\rm F}$ and $\kappa^2 = l^{-2}_{\rm coh} + \alpha k^2_{\perp} + \beta \cos(k_{\parallel}) - k^2_{\rm F}$, and $N_{{\bm k}_{\rm s}}$ is the normalization constant. }

{Following Ref.~\cite{wu21}, we expand $\Sigma^{\rm ss}_{ij}({\bm k}_{\rm s})$ in terms of $k_{\perp}$ and split the kernel into two terms, $K_{\mu\nu}^{\rm para}=K_{\mu\nu}^{(0)}+K_{\mu\nu}^{(2)}$, where
\begin{align}
    K_{\mu\nu,ij}^{(0)} \propto  & 
    \int_S \frac{d^2{\bm k}_{\rm s}}{(2\pi)^2} 
    J_{\mu}({\bm k}_{\rm s})J_{\nu}({\bm k}_{\rm s})
    \frac{\partial f(E_{{\bm k}_{\rm s}})}{\partial E_{{\bm k}_{\rm s}}}
    \Sigma^{\rm ss}_{ij}(k_{\perp}=0,k_{\parallel}), \\
    K_{\mu\nu,ij}^{(2)} \propto & 
    \int_S \frac{d^2{\bm k}_{\rm s}}{(2\pi)^2} 
    J_{\mu}({\bm k}_{\rm s})J_{\nu}({\bm k}_{\rm s})
    \frac{\partial f(E_{{\bm k}_{\rm s}})}{\partial E_{{\bm k}_{\rm s}}} %\nn \\
    \frac{\partial^2\Sigma^{\rm ss}_{ij}(0,k_{\parallel})}{\partial^2k_{\perp}}k^2_{\perp},
    \label{eq:K2}
\end{align}
The ${\bm k}_{\rm s}$-integration runs over the momentum region where the surface states exist, i.e., $k_{\perp}\in [-g(k_{\parallel}),g(k_{\parallel})]$ and $k_{\parallel}\in [-k_{{\rm F},\parallel},k_{{\rm F},\parallel}]$. Here, $k_{{\rm F},\parallel}$ satisfies $\epsilon_0(k_{\perp}=0,k_{\parallel}=k_{{\rm F},\parallel})=0$, and the Fermi momentum at $k_{\parallel}$ is defined as $g^2(k_{\parallel})=[k^2_{\rm F}-\beta\cos(k_{\parallel})]/\alpha$. The Majorana arcs have endpoints when $k_{{\rm F},{\parallel}}$ is defined within the surface Brillouin zone.}

{Let us first consider the screening current response flowing to the $k_{\perp}$ direction, i.e., the dispersive direction of the Majorana arc. The dominant low-temperature power law originates from $K_{\mu\nu}^{(0)}$ with $J_{\mu}=J_{\nu} \propto k_{\perp}$. By performing the $k_{\perp}$ integration, the kernel reduces to 
\begin{align}
    K_{\perp,ij}\approx K_{\perp,ij}^{(0)} \sim T^2,
\end{align}
for $T\rightarrow 0$. The $T^2$ dependence is rather robust and does not rely on the presence of the Majorana arc endpoints. After the $k_{\perp}$ integration in $K_{\mu\nu}^{(0)}$, the temperature dependence arises essentially from the factor $k_{\perp}^2(\partial f/\partial E)$ associated with the surface dispersion $E_s=-\Delta k_{\perp}$. Since the derivative of the Fermi function restricts the integral to $|E_s|\sim T$, one has $|k_{\perp}|\sim T$, and the resulting contribution scales as $k_{\perp}^2\sim T^2$. The remaining $k_{\parallel}$ dependence enters only through the smooth function $\Sigma^{ss}_{ij}(0,k_{\parallel})$ and therefore does not modify the power law. Consequently, even if the detailed shape of the Majorana arc is altered—for instance, by removing the arc endpoints—the leading temperature dependence remains proportional to $T^2$.}

{We now move on to the current response along the $k_{\parallel}$ direction. After the $k_{\parallel}$ integration, the contribution $K_{\parallel,ij}^{(0)}$ can be expressed as
\begin{align}
K_{\parallel,ij}^{(0)} \propto \int^{k_{{\rm F},\parallel}}_{0} dk_{\parallel}
A_{ij}(k_{\parallel})
\tanh\left(\frac{\tilde{E}_{k_{\parallel}}}{2T}\right),
\end{align}
where $A_{ij}(k_{\parallel})\propto \sin^2(k_{\parallel})\Sigma^{\rm ss}_{ij}(0,k_{\parallel})$. Importantly, $A_{ij}(k_{\parallel})$ is a smooth function of $k_{\parallel}$ and does not introduce additional temperature dependence. We also note that the characteristic energy scale,
\begin{align}
\tilde{E}(k_{\parallel})=\frac{\Delta}{k_{\rm F}} \sqrt{\frac{k_{\rm F}^2-\beta\cos(k_{\parallel})}{\alpha}},
\label{eq:Etilde}
\end{align}
appears after the integration of the kernel $K^{(0)}_{\parallel,ij}$ over $k_{\perp}$. We will show below that the energy scale plays a key role in determining the temperature dependence of the kernel. Expanding the hyperbolic tangent, $\tanh(x) =1-2e^{-2x}+O(e^{-4x})$, one sees that the first term gives a temperature-independent contribution, while the temperature dependence arises from
\begin{align}
K_{\parallel,ij}^{(0)}
 \propto {\rm const.} + C
\int^{k_{{\rm F},\parallel}}_{0} dk_{\parallel} A_{ij}(k_{\parallel})e^{-\tilde{E}(k_{\parallel})/T}.
\label{eq:K0}
\end{align}
where $C$ is the $T$-independent coefficient.} 

{Let us first consider the case where the Majorana arc has endpoints at $k_{\parallel}=\pm k_{{\rm F},\parallel}$, which are defined as $\tilde{E}(k_{{\rm F},\parallel})=0$. In the low temperature limit, $T\rightarrow 0$, the exponential factor, $e^{-\tilde{E}_{k_{\parallel}}/T}$, in $K^{(0)}_{\parallel,ij}$ strongly restricts the low-energy phase space to regions close to the arc endpoints. To understand the significant suppression of $K^{(0)}_{\parallel,ij}$ in the presence of the arc endpoints, it is useful to perform a simple power-counting analysis. Close to the arc endpoints, the surface dispersion behaves as $\tilde{E}(k_{\parallel})\propto\delta k_{\parallel}$, where $k_{\parallel}$ is the distance from the endpoint. The thermally active region is determined by $E(k_z)\sim T$, implying $\delta k_{\parallel}\sim T^2$. The function $\Sigma_{ss}(0,k_z)$ in Eq.~\eqref{eq:K0} is proportional to $\delta k_{\parallel}^2$, so that within the thermally relevant region it contributes a factor of order $T^4$. By contrast, the current vertex varies smoothly, resulting in a temperature-independent constant. Combining these factors shows that the contribution from the endpoint region is strongly suppressed at low temperatures, leading to the higher power $T^6$. The second contribution, $K_{\parallel,ij}^{(2)}\propto T^2$, follows directly from the $k_{\perp}^2(-\partial_E f)$ factor in Eq.~\eqref{eq:K2} and is largely insensitive to the detailed structure of the Majorana arc~\cite{wu21}.}

{We now consider a situation in which the Majorana arc extends smoothly across the surface Brillouin zone and no arc endpoints are present. The characteristic energy in Eq.~\eqref{eq:Etilde} has a finite gap, $\tilde{E}_{\min}=\min_{k_{\parallel}}\tilde{E}(k_{\parallel})\neq 0$, for $\forall{k_{\parallel}}$. In this case, both $A_{ij}(k_{\parallel})$ and $\tilde{E}(k_{\parallel})$ remain regular functions of momentum, and no special momentum point controls the $k_{\parallel}$ integration. The resulting temperature dependence of the kernel is given by 
\begin{align}
K_{\parallel,ij}^{\rm para} \sim {\rm const.} + a e^{-\tilde{E}_{\rm min}/T} + b T^2,
\end{align}
where $a$ and $b$ are $T$-independent constants. 
At temperatures well below this scale,
$T\ll \tilde{E}_{\min}$,
the exponential factor suppresses the contribution from $K^{(0)}_{\parallel,ij}$,
so that the total response is dominated by
\begin{align}
K_{\parallel,ij}^{\rm para} \approx K_{\parallel,ij}^{(2)} \sim T^2.
\end{align}
When the temperature becomes comparable to $\tilde{E}_{\min}$, $T\sim \tilde{E}_{\rm min}$ 
the rapid growth of $K^{(0)}_{\parallel,ij}\sim e^{-\tilde{E}_{\rm min}/T} $ increases the apparent power, producing an intermediate regime that can be phenomenologically described as
\begin{align}
K^{\rm para}_{\parallel,ij}\sim T^{2+\eta(T)}.
\end{align}
Since this enhancement originates from a crossover rather than a singularity of the dispersion, the additional exponent, $\eta(T)$, is not universal and depends on the detailed form of $\tilde{E}(k_z)$ and $T$.}

\bibliography{paper}

\end{document}